\documentclass[11pt,a4paper]{article}
\pdfoutput=1
\usepackage{jheppub}
\usepackage{gensymb}
\usepackage{placeins}
\usepackage{subfigure}
\usepackage{amssymb,amsmath}
\usepackage{graphicx}
\usepackage{color}
\usepackage{cancel}
\usepackage[colorlinks=true
,urlcolor=blue
,citecolor=blue
,linkcolor=blue
,pagecolor=blue
,linktocpage=true
,pdfproducer=medialab
]{hyperref}

 \usepackage[numbers]{natbib}
\usepackage{notoccite}
\makeatletter \renewcommand{\@dotsep}{10000} \makeatother
\def\be{\begin{equation}}
\def\ee{\end{equation}}
\def\bea{\begin{eqnarray}}
\def\eea{\end{eqnarray}}
\def\bi{\begin{itemize}}
\def\ei{\end{itemize}}

\newcommand{\cred}[1]{{\bf \color{red} #1}}
\newcommand{\cblue}[1]{{\bf \color{blue} #1}}
\newcommand{\WMSSM}{W_{{\rm MSSM}}}
\usepackage[nodisplayskipstretch]{setspace}

\newcommand{\mgut}{M_{{\rm GUT}}}
\newcommand{\msusy}{M_{{\rm SUSY}}}
\newcommand{\FTEW}{\Delta_{{\rm EW}}}
\newcommand{\Rtbtau}{R_{tb\tau}}

\DeclareUnicodeCharacter{2212}{-}
\DeclareUnicodeCharacter{202F}{\,}
\DeclareUnicodeCharacter{2032}{\ensuremath{-}}

\begin{document}

\begin{titlepage}
\pagestyle{empty}

\vspace*{0.2in}
\begin{center}
{\Large \bf Low Fine-Tuning with Heavy Higgsinos in Yukawa Unified SUSY GUTs} \\
\vspace{1cm}
{\bf Cem Salih $\ddot{\rm U}$n\footnote{E-mail: cemsalihun@uludag.edu.tr}}
\vspace{0.5cm}

{\small \it Department of Physics, Bursa Uluda\~{g} University, TR16059 Bursa, Turkey}
\end{center}

\vspace{0.5cm}
\begin{abstract}

The work presented considers a class of minimally constructed Yukawa unified SUSY GUTs - NUHM2 - and explore their implications when their soft supersymmetry breaking Lagrangian is generalized by the non-holomorphic terms which provide extra contributions to the Higgsino mass and couple the supersymmetric scalar fields to the wrong Higgs doublets. With such a simple extension, it can be found several regions with interesting implications which cannot be realized in the usual restricted models. It is observed that the Yukawa unification solutions can be compatible with relatively light mass spectrum and acceptable low fine-tuning measurements. In the restricted models such effects can directly be addressed to the non-holomorphic terms. They can provide a slight improvement in the SM-like Higgs boson mass without altering the mass spectrum too much, and they can accommodate relatively lighter sbottom and stau masses, while they do not change the stop sector much. The dark matter can be Higgsino-like or Bino-like, but the experimental relic density measurements favor the Higgsino-like dark matter, while the Bino-like dark matter is predicted with a quite large relic density. Also several coannihilation scenarios are identified in the Higgsino-like dark matter regions, while the Bino-like dark matter do not allow any of such coannihilation processes. The presence of the non-holomorphic terms can weaken the impact from the phenomenological or indirect constraints such as low fine-tuning, Yukawa unification and rare decays of $B-$meson, the direct and model independent constraints still yield a strong strike on the solutions. Such constraints are discussed in regard of the current collider analyses on $\tau\tau$ events and direct detection of dark matter experiments.

\end{abstract}
\end{titlepage}


\section{Introduction}
\label{sec:intro}

Despite its enormous success in explaining the observed phenomena in the particle physics, the Standard Model (SM) is assumed to only be an effective theory due to its drawbacks, which have been motivating the studies for new physics beyond the SM (BSM). Among uncountable candidates, supersymmetry (SUSY) is one of the forefront candidates for the BSM physics by proposing solutions to some of the profound problems of the SM such as the gauge hierarchy problem \cite{Gildener:1976ai,Gildener:1979dd,Weinberg:1978ym,Susskind:1978ms,Veltman:1980mj}, absence of a suitable dark matter (DM) candidate etc. In addition, the unification of the SM gauge couplings at a grand unification scale ($\mgut \simeq 2\times 10^{16}$ GeV) within the minimal supersymmetric SM (MSSM) allows one to explore the testable implications of SUSY grand unified theories (GUTs). Even though formulating the new physics by employing SUSY is well motivated for various theoretical reasons mentioned above, the absence of a direct signal for the supersymmetric particles weakens the strength the motivations behind the SUSY models. A straightforward conclusion, which can be derived from the null experimental results, is to shift the supersymmetric particle masses to the heavy scales. However, this approach can bring some problems of the SM back. For instance, if the mass difference between top-quarks and its superpartner stop is quite large, it can lead to large quadratic contributions to the Higgs boson mass which disturb the mass stability of the Higgs boson. Even if one can maintain the stability of the Higgs boson, then the $\mu-$term, which is expected to be at the order of electroweak symmetry breaking scale ($\mathcal{O}(100)$ GeV), can be realized at a few TeV scale and it leads to little hierarchy problem \cite{Gogoladze:2012yf,Gogoladze:2013wva}. 

In another approach, the lack of the direct signal can be considered as a missing part in MSSM. Such a missing part can be filled by extending the particle content of the MSSM  (for a recent studies see, for instance, \cite{Domingo:2022pde,Bagnaschi:2022zvd,Almarashi:2022iol,Biekotter:2021qbc}), non-universal boundary conditions induced by the symmetry breaking at $\mgut$ \cite{Lazarides:2023iim,Antusch:2023zjk}, higher order operators and/or representations \cite{Martin:2009ad,Chattopadhyay:2009fr,Ananthanarayan:2007fj} or extending the gauge group of MSSM \cite{Athron:2015tsa,Hicyilmaz:2016kty,Athron:2015vxg,Athron:2016gor,Hicyilmaz:2017nzo,Frank:2021nkq,Hicyilmaz:2021khy}. Apart from such possible paths, one of relatively simpler ways to extend MSSM is to generalize its soft supersymmetry breaking (SSB) Lagrangian by considering the non-holomorphic (NH) terms. In supersymmetrizing the SM and BSM models, it has been very well known that the superpotential cannot be a functional of a field and its complex conjugate simultaneously \cite{Martin:1997ns}, which is called holomorphy condition. Even though SSB can induce such NH terms \cite{Jack:1999ud,Jack:1999fa}, most of the earlier works respected the holomorphy condition even after SSB. Despite being a simple extension, NH can provide a more general construction of MSSM, and they can significantly alter the implications of MSSM such as the mass spectrum of the supersymmetric particles (sparticles) and Higgs bosons, the rare semi-leptonic rare $B-$meson decays, the dark matter implications \cite{Un:2014afa,Rehman:2022ydc,Chattopadhyay:2022ecq,Chattopadhyay:2018tqv}.

This work considers the impact from such NH terms in frameworks of SUSY GUTs. Among a large variety of SUSY GUT models, those based on $SU(5)$ \cite{Chattopadhyay:2001mj,Komine:2001rm,Profumo:2003ema,Pallis:2003aw,Balazs:2003mm,Altmannshofer:2008vr,Gogoladze:2008dk,Antusch:2009gu,Baer:2012by,Gogoladze:2010fu} and $SO(10)$ \cite{Rattazzi:1995gk,Blazek:1996yv,Baer:1999mc,Gomez:1999dk,Gogoladze:2003pp,Chattopadhyay:2001va,Blazek:2001sb,Baer:2009ie} have been attracting the attention and excessively explored in their implications testable in the collider and DM experiments. These GUT models are also interesting, since they do not unify only the gauge couplings, but in their minimal constructions they propose some schemes of Yukawa Unification (YU) such as $b-\tau$ YU in $SU(5)$ and $t-b-\tau$ YU in $SO(10)$ GUTs. Despite being imposed at the GUT scale, the YU scenarios lead to a significant impact on the low scale implications and make the GUT models interestingly predictive. This behavior of YU arises from the need for the threshold corrections at the SUSY scale ($\msusy$) at which the supersymmetric particles are assumed to be decoupled from the spectrum. These threshold corrections play a crucial role to realize YU consistent with the third family fermion masses (see, for instance, \cite{Gogoladze:2010fu,Shafi:2023sqa}).

In addition to the threshold corrections, YU is also a strong condition which significantly shapes the fundamental parameter space. The large mass difference between the top and bottom quarks, for instance, happens to approximately equal to $\tan\beta$, thence YU can be realized when $\tan\beta \gtrsim m_{t}/m_{b} \simeq 40$, where $\tan\beta$ is defined as the ratio of the vacuum expectation values (VEVs) of the two Higgs doublets of MSSM as $tan\beta \equiv v_{u}/v_{d}$.  Another impact arises from the radiative electroweak symmetry breaking (REWSB). One of the requirements to realize REWSB consistently leads to $m_{H_{u}} < 0$ and $m_{H_{u}} < m_{H_{d}}$, where $m_{H_{u}}$ and $m_{H_{d}}$ are the SSB masses of the MSSM Higgs doublets. If one imposes the YU condition at the GUT scale, a solution consistent with REWSB can be realized only when $m_{H_{u}}\neq m_{H_{d}}$ at $\mgut$ \cite{Olechowski:1994gm,Matalliotakis:1994ft,Murayama:1995fn,Baer:2009ie}, especially when the gaugino masses are set universal at the GUT scale. Hence, the models with Non-Universal Higgs Masses Type 2 (NUHM2 with $m_{H_{d}}\neq m_{H_{u}}$) form the minimal construction for the Yukawa unified SUSY GUTs. In this context, despite its restrictive impact, the YU condition imposed at $\mgut$ yields quite predictive SUSY GUT models \cite{Gogoladze:2011aa,AdeelAjaib:2013dnf} in terms of the SM-like Higgs boson mass, supersymmetric particle spectrum and DM implications.

Considering the motivations mentioned above, it might be worth exploring the implications of the SUSY GUTs in the presence of NH terms. The minimally constructed models in the class of NUHM2 is taken under consideration in this work to make the effects and the implications of NH terms more visible for the SM-like Higgs boson mass, Yukawa unification and supersymmetric mass spectrum as well as the predictions for DM in terms of its composition, relic density and scattering at matter nuclei which can be tested in direct detection experiments of DM. Even though it would not be realistic to remove the tension between the latest experimental results and the constrained SUSY GUTs, this tension can be alleviated by the NH terms. In addition to such observable parameters, the presence of NH terms can yield low fine-tuned SUSY GUT models. The fine-tuning measurements can be interpreted as the effectiveness of the missing parts, and usually the restricted SUSY GUTs lead to a significantly large fine-tuning \cite{Baer:2012mv,Demir:2014jqa}. Thus, the low fine-tuning results can mean that the presence of NH terms can yield a considerable compensations in the effects of the missing parts in SUSY models. This approach is not restricted with MSSM, and it can be applied to the extended SUSY models as well \cite{Demir:2005ti,Hayreter:2007it}.

The rest of the paper is organized as follows: It will be discussed the inclusion of the NH terms and their contributions to the model implications in Section \ref{sec:NHLev}. After briefly describing the scanning procedure and the experimental constraints employed in the analyses in Section \ref{sec:scan}, it will be discussed the mass spectrum for the Higgs bosons and supersymmetric particles in Sections \ref{sec:FTmu} and \ref{sec:spectrum}. Section \ref{sec:DM} will focus on the DM implications in terms of the DM composition as well as the observable DM parameters along with the current experimental results. This section also has a table of benchmark points which exemplify our results. The findings will be summarized and concluded in Section \ref{sec:Conc}.

\section{Leverage for NUHM2 from NH Terms}
\label{sec:NHLev}

Before the SUSY breaking, MSSM can be constructed through the following superpotential 

\begin{equation}
\WMSSM = y_{u}QH_{u}\bar{u} + y_{d}QH_{d}\bar{d} + y_{e}LH_{d}\bar{e} + \mu H_{u}H_{d}~,
\label{eq:WMSSM}
\end{equation}
where $Q$, $L$, $u$, $d$, $e$ represent the superfields for the left-handed quarks, left-handed leptons, right-handed up-quark, right-handed down quark and right-handed lepton, respectively. the family indices are suppressed but $y_{u}$, $y_{d}$ and $y_{e}$ are $3\times 3$ Yukawa matrices for up-type, down-type quarks and leptons, respectively. Note that due to the holomorphy condition, one Higgs doublet is not enough to drive the masses for all the fields, thus one needs to employ a second Higgs doublet in Eq.(\ref{eq:WMSSM}), which are distinguished in the notation by $H_{u}$ and $H_{d}$. The superpotential also has a bilinear mixing term for these Higgs doublets quantified by the $\mu-$term. This $\mu-$term is of a special importance, since it is directly relevant to the electroweak symmetry breaking, fine-tuning measurement and the Higgsino masses, which will be discussed below in details. 

Despite the simple form of the superpotential, the SSB drives the scalar potential to a more complicated form by inducing the mass terms for the supersymmetric partners for the SM fermions (sfermions), gauge bosons (gauginos) and the Higgs doublets. In addition, the SSB Lagrangian include Yukawa interactions between the MSSM Higgs fields and sfermions characterized by dimensionful couplings denoted as $A_{u,d,e}$, which are formed by $3\times 3$ matrices as well. However, in the usual treatment, the sfermions are kept interacting with the appropriate Higgs doublets as fixed in the superpotential such that the holomorphy condition holds for the SSB Lagrangian as well. This assumption can still provide a good approximation in exploring the implications of SUSY models, since most of NH terms are suppressed by $M_{Z}/M$ where $M_{Z}$ represents the low scale which is identified as the $Z-$boson mass, and $M$ denotes the high energy scale at which the symmetry breaks into the MSSM gauge group. Such a suppression can be realized when the SUSY is broken by VEVs of gauge singlets, while in other cases the following terms can still have significant magnitudes \cite{Jack:1999fa}

\begin{equation}
-{\cal{L}}_{soft}^{\prime} =\mu^\prime {\tilde H_u}\cdot
{\tilde H_d} +\tilde{Q}~{H}_d^{\dagger} A^\prime_{u} \tilde{U}+
\tilde{Q}~ {H}_u^{\dagger} A^\prime_{d} \tilde{D}
 + \tilde{L}~{H}_u^{\dagger} A^\prime_{e} \tilde{E} + \mbox{h.c.}~,
\label{eq:nonholSSB}
\end{equation}
where $\mu^{\prime}$ is the Higgsino mixing term, and it contributes only to the Higgsino mass, while the radiative electroweak symmetry breaking (REWSB) remains only dependent on $\mu-$term \cite{Dedes:1995sb}. The rest of the terms with $A_{u,d,e}^{\prime}$ are the NH correspondence of the trilinear scalar interaction terms of the holomorphic SSB Lagrangian. Their NH correspondence arises the interactions of sfermions with the wrong Higgs doublet of MSSM. Even such a simple extension of MSSM can yield considerable impacts, which will be discussed in separate subsections:

\subsection{Low Fine-Tuning and Higgsino Mass}
\label{subsec:FTHiggsinoMass}

The Fine-tuning in SUSY models can be measured by employing the following equation arising from the minimization of the scalar Higgs potential after the electroweak symmetry breaking:

\begin{equation}
\dfrac{M_{Z}^{2}}{2}=-\mu^{2}+\dfrac{m_{H_{d}}^{2}-m_{H_{u}}^{2}\tan^{2}\beta}{\tan^{2}\beta -1}.
\label{eq:MZ}
\end{equation}
where SSB masses for the Higgs doublets, $m_{H_{d}}$ and $m_{H_{u}}$, are substituted after the loop contributions are added to them. Following the usual definition in quantifying the fine-tuning \cite{Baer:2012mv} measure one can write

\begin{equation}
\FTEW\equiv \dfrac{{\rm Max}(C_{i})}{M_{Z}^{2}/2}
\label{eq:delfine}
\end{equation}
with

\begin{equation}
\setstretch{1.5}
C_{i}=\left\lbrace
\begin{array}{l}
C_{H_{d}}=\mid m^{2}_{H_{d}}/(\tan^{2}\beta -1) \mid~, \\ 
C_{H_{u}}=\mid m^{2}_{H_{u}}\tan^{2}\beta/(\tan^{2}\beta -1) \mid~, \\ 
C_{\mu}=\mid -\mu^{2}\mid~.
\end{array}
\right.
\label{eq:CFT}
\end{equation}

In the absence of NH terms, the fine-tuning is usually quantified with $C_{\mu}$. Indeed, $C_{H_{d}}$ is suppressed by $\tan\beta$, and a consistent measure of $M_{Z}$ requires $C_{H_{u}} \approx C_{\mu}$. If one applies the condition $\FTEW \leq 100$ (following \cite{Gogoladze:2012yf,Li:2016ucz,Baer:2018rhs}) to identify the solutions with acceptable fine-tuning, this condition yields $\mu\lesssim 600$ GeV. When the NH terms are absent, the acceptable fine-tuning condition also determines the Higgsino masses, since their masses are equal to $\mu$. In other words, in the acceptably fine-tuned regions of MSSM, one should expect the Higgsino masses to be around 600 GeV or lighter. When the Higssino is the lightest supersymmetric particle (LSP), which is assumed to be stable, they can also be considered a DM candidate. However, if the DM is formed by such light Higgsinos, solutions receive a strong impact from the latest results of the Planck satellite \cite{Planck:2018nkj}, which requires $\mu \gtrsim 700$ GeV. This bound on the LSP Higgsino mass is shifted further up to about 1 TeV \cite{Raza:2018jnh,Hicyilmaz:2023tnr,Shafi:2023sqa} by the current results from the direct DM detection experiments \cite{Akerib:2018lyp,Aalbers:2016jon,Aprile:2020vtw,Tanaka:2011uf,Khachatryan:2014rra,Abbasi:2009uz,Akerib:2016lao}. Hence, one needs to consider possible extensions in MSSM to fit the low fine-tuning solutions with the experimental results on DM observables \cite{Evans:2022gom,Baer:2018rhs}. Even though the bounds do not apply directly to Higgsino when the DM is formed by one of the MSSM gauginos or their mixing with Higgsinos, these experimental results still yield a strong impact on the Higgsino mass and consequently on $\FTEW$. This tension between the experimental results on the Higgsino DM and the low fine-tuned solutions can be removed by extending MSSM with the NH terms, which drive the Higgsino mass as $m_{\tilde{H}} \simeq \mu + \mu^{\prime}$, while they leave Eq.(\ref{eq:MZ}) intact. In this case one can still realize consistently heavy Higgsinos, while the fine-tuning remains low (i.e. $\mu^{\prime} \gg \mu$). The interplay between the DM implications and the relevant experimental constraints will be discussed in Section \ref{sec:DM} in detail.

\subsection{SUSY Mass Spectrum}
\label{subsec:SUSYSpectra}

In addition to the low fine-tuning and consistent Higgsino DM solutions, the NH trilinear interaction couplings ($A_{\tilde{f}}^{\prime}$, where $\tilde{f} = \tilde{u},\tilde{d},\tilde{e}$) in Eq.(\ref{eq:nonholSSB}) can be effective in the SUSY spectra. The NH contributions can be seen from the mass-square matrix of the sfermions given as \cite{Jack:1999ud,Jack:1999fa}: 

\begin{equation}
\setstretch{1.5}
m_{\tilde{f}}^{2} = \left( \begin{array}{cc}
m_{\tilde{f}_{L}\tilde{f}^{*}_{L}} & X_{\tilde{f}} \\
X_{\tilde{f}}^{*} & m_{\tilde{f}_{R}\tilde{f}^{*}_{R}}
\end{array} \right)~,
\label{sfermionsmass2}
\end{equation}
where the subindex $\tilde{f} = \tilde{u},\tilde{d},\tilde{e}$ denotes the up-type squarks, down-type squarks and sleptons, respectively. The form of $m_{\tilde{f}}^{2}$ is typical and the diagonal components of $m_{\tilde{f}}^{2}$ are the same as those in MSSM \cite{Martin:1997ns}, the NH terms appear in the off-diagonal terms ($X_{\tilde{f}}$) and they can change the mixing and the mass eigenvalues of the sparticles significantly. These terms can be summarized for each field as follows:

\begin{equation*}
X_{\tilde{u}} = -\frac{1}{\sqrt{2}}[v_{d}(\mu Y^{\dagger}_{u}+A^{'\dagger}_{u})-v_{u}A_{u}^{\dagger}] \tag{2.6-a}
\label{eq:Xu}
\end{equation*}
\begin{equation*}
X_{\tilde{d}} = -\frac{1}{\sqrt{2}}[v_{u}(\mu Y_{d}^{\dagger}+A^{'\dagger}_{d})-v_{d}A_{d}^{\dagger}] \tag{2.6-b}
\label{eq:Xd}
\end{equation*}
\begin{equation*}
X_{\tilde{e}} = \frac{1}{\sqrt{2}}[-v_{u}(\mu Y_{e}^{\dagger}+A^{'\dagger}_{e})+v_{d}A_{e}^{\dagger}] \tag{2.6-c}
\label{eq:Xe}
\end{equation*}
\setcounter{equation}{6}

As seen from Eq.(\ref{eq:Xu}), the NH trilinear coupling in the stop (the superpartner of top quark) sector interferes in the stop mixing along with $v_{d}$, while in sbottom and stau given in Eqs.(\ref{eq:Xd} and \ref{eq:Xe}) they appear together with $v_{u}$. In this case, since $v_{d} \ll v_{u}$, the stop masses are not affected too much by the NH terms, unless $A_{t}^{\prime}$ is allowed to be very large. On the other hand, they can change significantly the sbottom and stau masses in the mass spectrum.  Even though it does not contribute to the masses of sfermions at tree-level, the $\mu^{\prime}-$term also alters the masses at loop-level considerably (see, for instance, \cite{Un:2014afa,Chattopadhyay:2018tqv}).

\subsection{Mass of Higgs Bosons}
\label{subsec:HiggsSpectrum}

Even though one can nicely fit the SM-like Higgs boson, one of the main strikes to MSSM comes from the experimental results for the Higgs boson mass \cite{ATLAS:2012yve,CMS:2013btf}. Since MSSM yields inconsistently light Higgs boson at tree-level, one needs to utilize the loop corrections to realize a consistent mass in the spectrum. Due to the negligibly small Yukawa couplings to the fermions of the first two families, only the third family fermions are relevant in calculations of the loop contributions. Even though the main contributions come from the stop sector, sbottom and stau can also  provide minor contributions together with moderate or large $\tan\beta$. The leading terms at loop level in the Higgs boson mass can be given as \cite{Carena:2012mw}

\begin{equation*}
\Delta m_{h}^{2}\simeq \dfrac{m_{t}^{4}}{16\pi^{2}v^{2}\sin^{2}\beta}\dfrac{\mu A_{t}}{M^{2}_{{\rm SUSY}}}\left[\dfrac{A_{t}^{2}}{M^{2}_{{\rm SUSY}}}-6 \right]+
\end{equation*}
\begin{equation}\hspace{1.4cm}
\dfrac{y_{b}^{4}v^{2}}{16\pi^{2}}\sin^{2}\beta\dfrac{\mu^{3}A_{b}}{m^{4}_{\tilde{b}}}+\dfrac{y_{\tau}^{4}v^{2}}{48\pi^{2}}\sin^{2}\beta \dfrac{\mu^{3}A_{\tau}}{m_{\tilde{\tau}}^{4}}~~,
\label{eq:higgscor}
\end{equation}
where the first line represents the contributions from stops, while those from the sbottom and stau are given in the second line. Even though the terms in the second line might be important, the stability of the Higgs vacuum severely constrains these contributions by bounding $\mu\tan\beta$ \cite{Carena:2012mw,Hisano:2010re,Kitahara:2013lfa}. However, in the presence of the NH terms, there will be more contributions added to those given in Eq.(\ref{eq:higgscor}) arising from $X_{b,\tau}$ \cite{Brignole:2002bz} defined in Eqs.(\ref{eq:Xd} and \ref{eq:Xe}), which are proportional to 

\begin{equation}
\Delta_{{\rm NH}}^{b,\tau} \propto \dfrac{y_{b,\tau}}{16\pi^{2}}\sin\beta^{2}\dfrac{\mu^{3}A_{b,\tau}^{\prime}}{m_{\tilde{b},\tilde{\tau}}^{4}}~.
\label{eq:NHcor1}
\end{equation}

The NH contributions can be realized from about 0.5 GeV to about 1 GeV depending on the size of the NH trilinear couplings, while these contributions can be even as large $\mathcal{O}(100)$ GeV for the heavy Higgs bosons of MSSM \cite{Un:2014afa,Rehman:2022ydc}.

Even though this section discussed mostly the direct impacts from the NH terms, they also lead some indirect effects as well. For instance, some solutions excluded by the rare semi-leptonic decays of $B-$meson can become available again \cite{Un:2014afa}, or removing the link between the Fine-tuning and Higgsino mass, one can realize available Higgsino DM solutions compatible with the acceptable fine-tuning solutions which are not observed in the absence of NH terms. In this context, even though it seems to be a simple extension of MSSM, the NH terms bring back to the game a significant portion of the parameter space. Thus, the analyses impose  an extra condition on the Yukawa couplings at the GUT scale to constrain the parameter space further. Hence, this study focuses on a class of SUSY models which propose YU at $\mgut$. As discussed in Section \ref{sec:intro}, YU yields another strong impact in shaping the parameter space of the models.

\section{Scanning Procedure and Experimental Constraints}
\label{sec:scan}

The fundamental parameter space is spurred by several free parameters which are formed mostly by the terms involved in the SSB Lagrangian. These terms and their ranges in our scans can be summarized as follows:

\begin{equation}
\setstretch{1.5}
\begin{array}{lll}
0 \leq & m_{0} & \leq 10 ~{\rm TeV} \\
0 \leq & M_{1/2} & \leq 10 ~{\rm TeV} \\
-3 \leq & A_{0}/m_{0} & \leq 3 \\
-3 \leq & A_{0}^{\prime}/m_{0} & \leq 3 \\
1.2 \leq & \tan\beta & \leq 60 \\
0 \leq & m_{H_{d}}, m_{H_{u}} & \leq 10 ~{\rm TeV}\\
0 \leq & \mu^{\prime} & \leq 5 ~{\rm TeV}\\
\end{array}
\label{eq:psranges}
\end{equation}

Even though the SSB Lagrangian has many more terms than those listed above, these terms cannot be arbitrary when they are constrained from $\mgut$. The underlying GUT scale gauge symmetry induces some relations among them. In our study NUHM2-like GUT models are assumed, which can arise from $SO(10)$ GUT groups. Following this pattern, all the matter fields of a family of SM can be fit in a single $\mathbf{16}$ dimensional representation of SO(10), thus SSB induces the same mass for all the fields of a family. In addition, one can assume a flavor universality at $\mgut$, thus all the three families can be assigned to the same mass, which is denoted by $m_{0}$ in Eq.(\ref{eq:psranges}). Similarly, if the GUT symmetry is formed by a single group, the three gauge boson of the SM emerges from the same adjoint representation of the GUT group, thus their super partners, gauginos, are assumed to have the same mass, and in our scans $M_{1/2}$ is assigned to all the MSSM gauginos at $\mgut$. $A_{0}$ and $A_{0}^{\prime}$ represent the holomorphic and NH trilinear scalar couplings, respectively. Instead of assigning a value directly to these parameters, they are varied with respect to their ratios to the supersymmetric scalar mass term in order to avoid the charge/color breaking vacua in the scalar potential \cite{Ellwanger:1999bv,Camargo-Molina:2013qva}. Since the matter fields are resided in the same representations, these couplings are set universal as $A_{u}=A_{d}=A_{e}=A_{0}$ and $A^{\prime}_{u}=A^{\prime}_{d}=A^{\prime}_{e}=A^{\prime}_{0}$. 

$\tan\beta$, $m_{H_{d}}$, $m_{H_{u}}$ and $\mu^{\prime}$ are the parameters which were defined in the previous section. Note that the $\mu-$term is not a free parameter in our scans and it is calculated in terms of the other free parameters through Eq.(\ref{eq:MZ}). However, this relation allows only to determine its magnitude while its sign remains arbitrary. In the scans, only the cases with positive $\mu$ ($\mu > 0$) are accepted.  Among the parameters listed in Eq.(\ref{eq:psranges}), the values for  all the parameters are imposed at $\mgut$, except $\mu^{\prime}$ which is varied at the low scale. 
 
Random scans are performed over these parameters by using SPheno \cite{Porod:2003um,Porod:2011nf} generated by SARAH \cite{Staub:2008uz,Staub:2015iza} to calculate the mass spectra, the cross-sections and the branching ratios for possible decay modes of the particles. SPheno calculates $\mgut$ dynamically by running renormalization group equations (RGEs) for the gauge couplings and employing the unification assumption as $g_{1}=g_{2}\simeq g_{3}$. After $\mgut$ is determined, the RGEs are run back to $M_{Z}$ together with the SSB parameters given in Eq.(\ref{eq:psranges}). The parameters varied at the low scale are substituted at $M_{Z}$. Even though SPheno employs advance methods in calculating the SM-like Higgs boson mass, there are about 3 GeV uncertainty arising from the uncertainties in the top-quark mass, {the QCD coupling} and the mixing in the stop sector \cite{Degrassi:2002fi}. The top quark mass is set to its central value ($m_{t}=173.3$ GeV \cite{CDF:2009pxd,ATLAS:2021urs}). Despite the insensitivity of the SUSY mass spectrum to the top quark mass, the SM-like Higgs boson mass can be enhanced 1-2 GeV by varying the top quark mass by $1-2\sigma$ in its experimental measurements \cite{AdeelAjaib:2013dnf,Gogoladze:2014hca}.

In the scanning procedure the Metropolis-Hastings algorithm \cite{Baer:2008jn,Belanger:2009ti} is followed, and the scans accept only the solutions in which the LSP happens one of the MSSM neutralinos (gauginos or higgsinos). With this selection, one can assume that LSP can account for the DM observables and the parameter space can be constrained further by the DM experiments. In this context the solutions are transferred from SPheno to micrOMEGAs \cite{Belanger:2018ccd} to add the dark matter (DM) observables in our analyses. After generating the data, the data is subjected to the mass bounds \cite{Agashe:2014kda}, constraints from combined results for rare $B-$meson decays \cite{CMS:2020rox,Belle-II:2022hys,HFLAV:2022pwe}, and the latest Planck Satellite measurements \cite{Planck:2018nkj} on the DM relic abundance successively to constrain the LSP neutralino. The list given below summarizes the experimental constraints employed in the analyses:
\begin{equation}
\setstretch{1.8}
\begin{array}{l}
m_h  = 123-127~{\rm GeV}\\
m_{\tilde{g}} \geq 2.1~{\rm TeV}~(800~{\rm GeV}~{\rm if~it~is~NLSP})\\
1.95\times 10^{-9} \leq{\rm BR}(B_s \rightarrow \mu^+ \mu^-) \leq 3.43 \times10^{-9} \;(2\sigma) \\
2.99 \times 10^{-4} \leq  {\rm BR}(B \rightarrow X_{s} \gamma)  \leq 3.87 \times 10^{-4} \; (2\sigma) \\
0.114 \leq \Omega_{{\rm CDM}}h^{2} \leq 0.126 \; (5\sigma)~.
\label{eq:constraints}
\end{array}
\end{equation}

In addition to these constraints the solutions with the acceptable fine-tuning are determined by applying the condition $\FTEW \leq 100$, where $\FTEW$ is calculated by following Eqs.(\ref{eq:delfine} and \ref{eq:CFT}). As mentioned above, the presence of NH terms brings back most of the previously excluded solutions to be experimentally consistent, thus a further condition on the Yukawa couplings that unifies them  is applied to constrain the parameter space. The YU condition can be parametrized with $R_{tb\tau}$ defined as follows:

\begin{equation}
R_{tb\tau} = \dfrac{{\rm max}(y_{t},y_{b},y_{\tau})}{{\rm min}(y_{t},y_{b},y_{\tau})}
\label{eq:RYU}
\end{equation}
where $y_{t}$, $y_{b}$ and $y_{\tau}$ represent the GUT scale values of Yukawa couplings of top-quark, bottom-quark and $\tau-$lepton respectively. In the ideal case in which $y_{t} = y_{b} = y_{\tau}$, $R_{tb\tau}$ will apparently be equal to 1, but solutions can be still considered to be compatible with YU if they satisfy the condition $R_{tb\tau} \leq 1.1$. This $10\%$ deviation is accounted for some unknown threshold corrections to the Yukawa couplings arising from the symmetry breaking at $\mgut$.

\section{Fine-Tuning, Higgsino Masses and Yukawa Unification}
\label{sec:FTmu}

\begin{figure}[t!]
\centering
\includegraphics[scale=0.4]{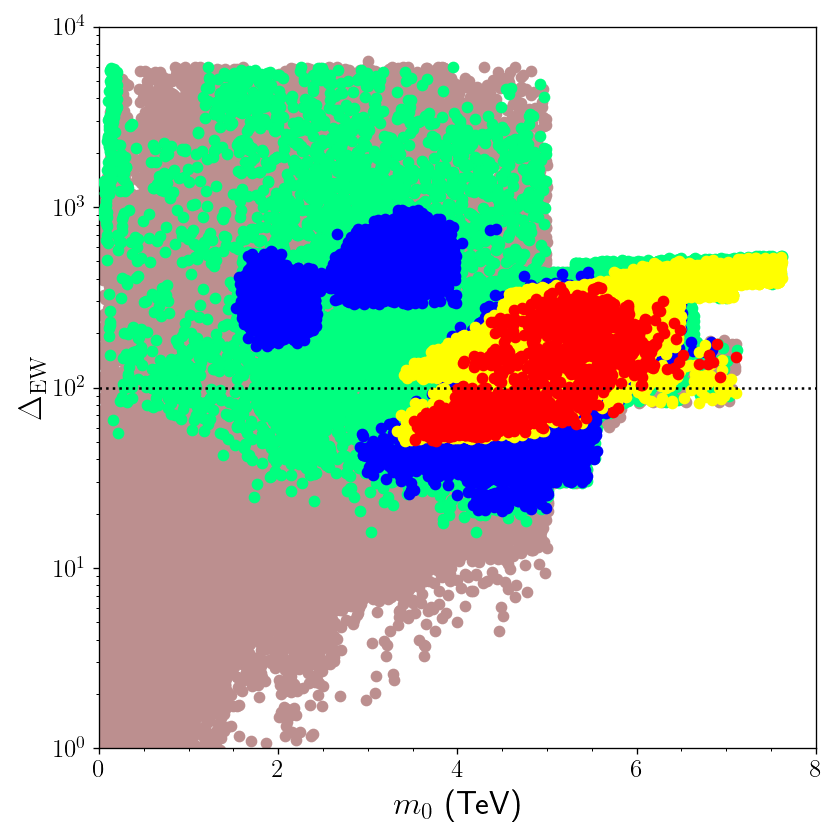}%
\includegraphics[scale=0.4]{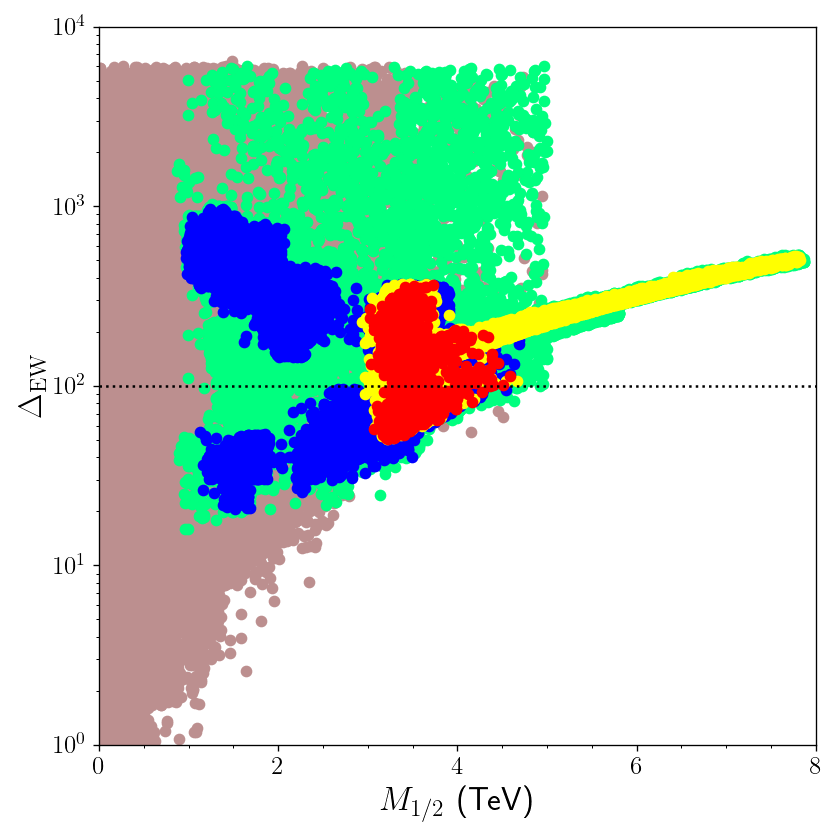}
\includegraphics[scale=0.4]{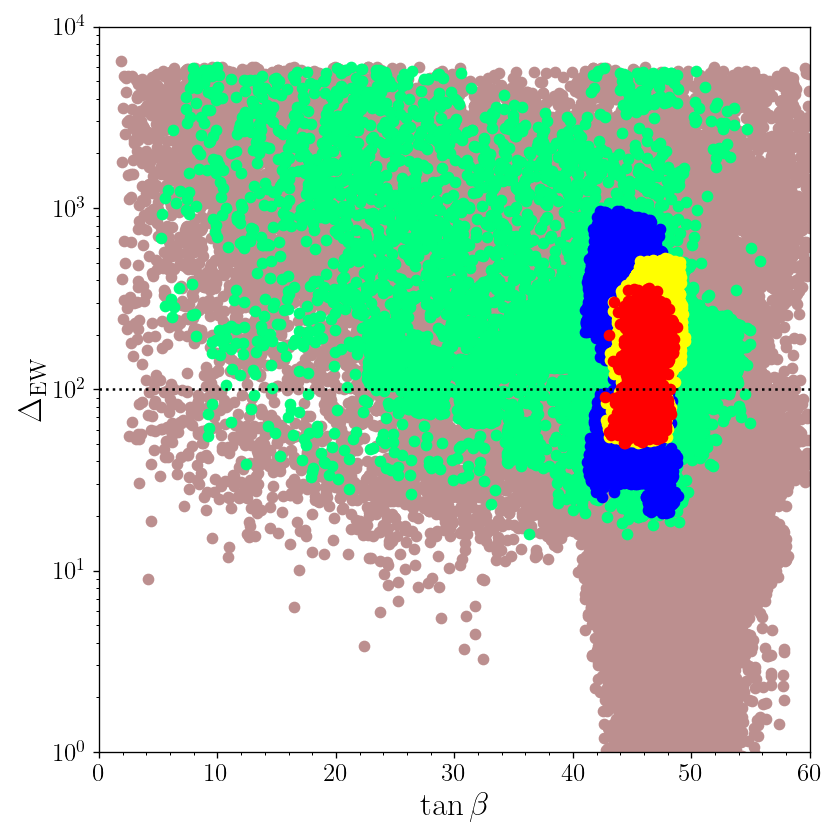}%
\includegraphics[scale=0.4]{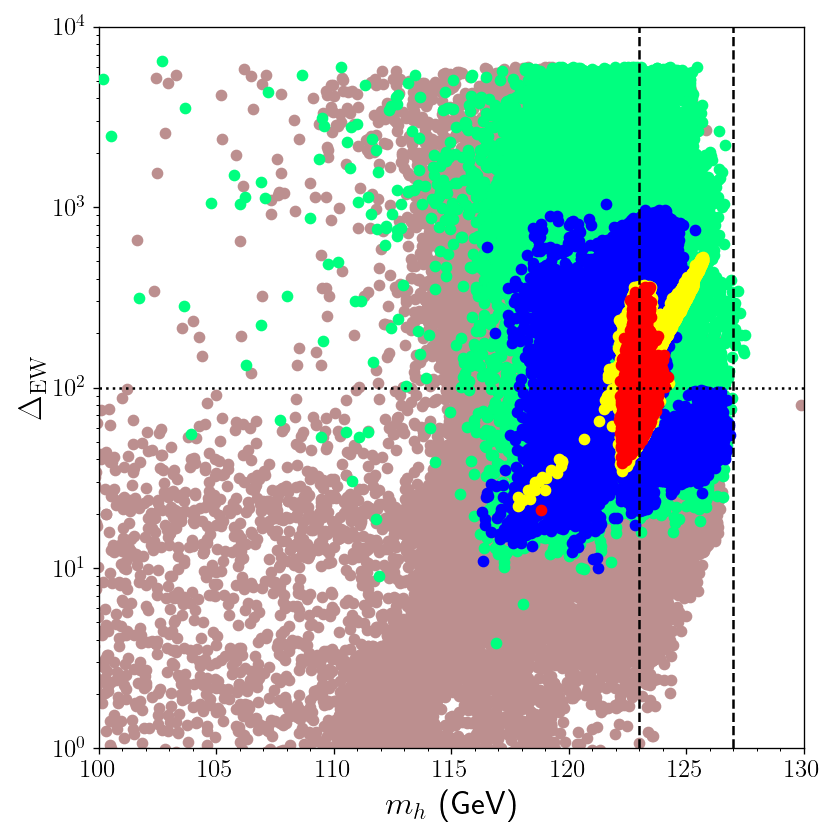}
\caption{The Fine-tuning plots in terms of the scalar and gaugino GUT masses (top), $\tan\beta$ and SM-like Higgs boson mass (bottom). All the points are compatible with REWSB and LSP neutralino condition. The green points are consistent with the mass bounds and constraints from rare $B-$meson decays. The blue points form a subset of green and they are identified as to be YU solutions. The red and yellow points form separate subsets of blue. The red points represent the solutions in which LSP neutralino relic density is compatible with the Planck measurements within $5\sigma$, while the yellow solutions indicate lower relic density of LSP neutralino. The Fine-tuning condition is not applied in these plots, while it is indicated horizontal dotted lines in the planes. In addition, in the bottom-right plane where the SM-like Higgs boson mass is shown, the Higgs boson mass constraint is not applied, but its range given in Eq.(\ref{eq:constraints}) are shown by the two vertical dashed lines in this plane.}
\label{fig:FTfundamental}
\end{figure}

One can first consider the impact from the fine-tuning measures in the fundamental parameters space together with the YU condition. Figure \ref{fig:FTfundamental} displays the magnitudes of $\FTEW$ in correlation with the fundamental parameters and the SM-like Higgs boson mass. All the points are compatible with REWSB and LSP neutralino conditions. The green points are consistent with the mass bounds and constraints from rare $B-$meson decays. The blue points form a subset of green and they are identified as to be YU solutions. The red and yellow points form separate subsets of blue. The red points represent the solutions in which LSP neutralino relic density is compatible with the Planck measurements within $5\sigma$, while the yellow solutions indicate lower relic density of LSP neutralino. The Fine-tuning condition is not applied in these plots, while it is indicated with horizontal dotted lines in the planes. In addition, in the bottom-right plane where the SM-like Higgs boson mass is plotted, the Higgs boson mass constraint is not applied, but its range given in Eq.(\ref{eq:constraints}) are shown by the two vertical dashed lines in this plane. The $\FTEW-m_{0}$ plane shows that the fine-tuning condition can bound the scalar masses at $\mgut$ from above at around 7 TeV (colored points), while the YU condition bounds it from below at about 3 TeV (blue points). These bounds become stronger for the SSB gaugino masses shown in the $\FTEW-M_{1/2}$ plane as to be $1 \lesssim M_{1/2} \lesssim 4.5$ TeV. Note that the lower bound on $M_{1/2}$ is as large as about 1 TeV due to the heavy mass bound on the gluino mass ($m_{\tilde{g}} \geq 2.1$ TeV). The strongest impact on $\tan\beta$ parameter arises from the YU condition, which can be realized when $\tan\beta \gtrsim 40$ (blue points), while the fine-tuning condition itself allows the solutions with $\tan\beta$ as low as about 10. 

The results shown in the $\FTEW-m_{h}$ plane can reveal the impact from the SM-like Higgs boson mass constraint. While the fine-tuning condition can favor inconsistently light Higgs boson ($m_{h} \gtrsim 105$ GeV), the YU condition shifts this bound up to about 115 GeV, which are already excluded experimentally. The consistent mass range for the Higgs boson is shown by the vertical dashed lines, and despite its strong negative impact, the fine-tuning and YU condition can easily fit with the consistent Higgs boson mass. Indeed, the main bound in this region comes from the DM observables. The Planck measurements (red points) can bound the SM-like Higgs boson mass at about 124.5 GeV from below, while the lower relic density solutions (yellow) allow the Higgs boson mass at about 126 GeV in the mass spectrum.

\begin{figure}[t!]
\centering
\includegraphics[scale=0.4]{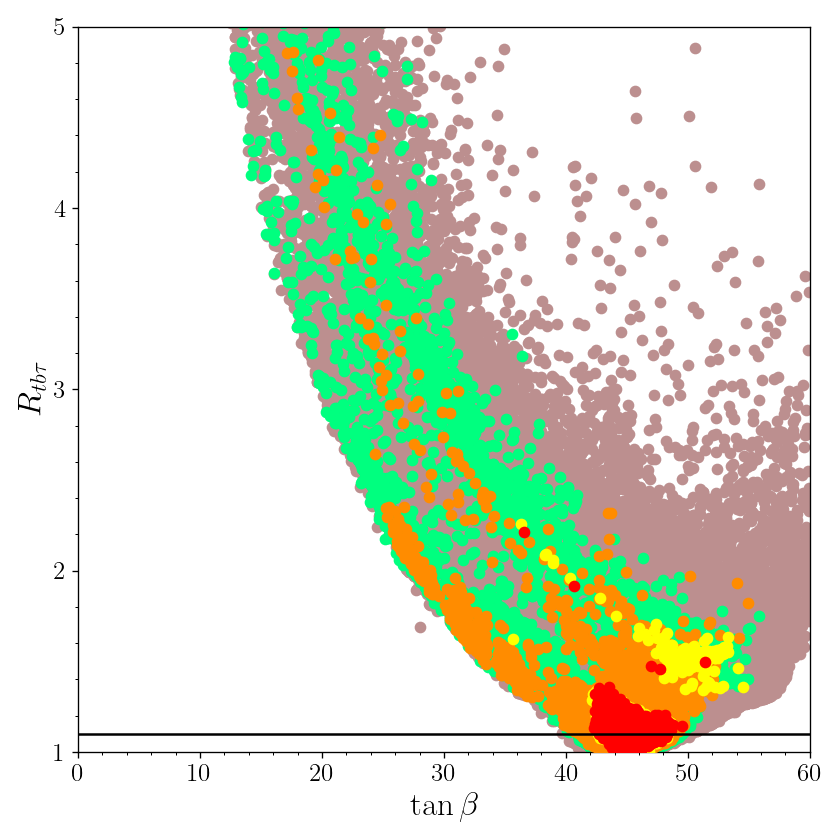}%
\includegraphics[scale=0.4]{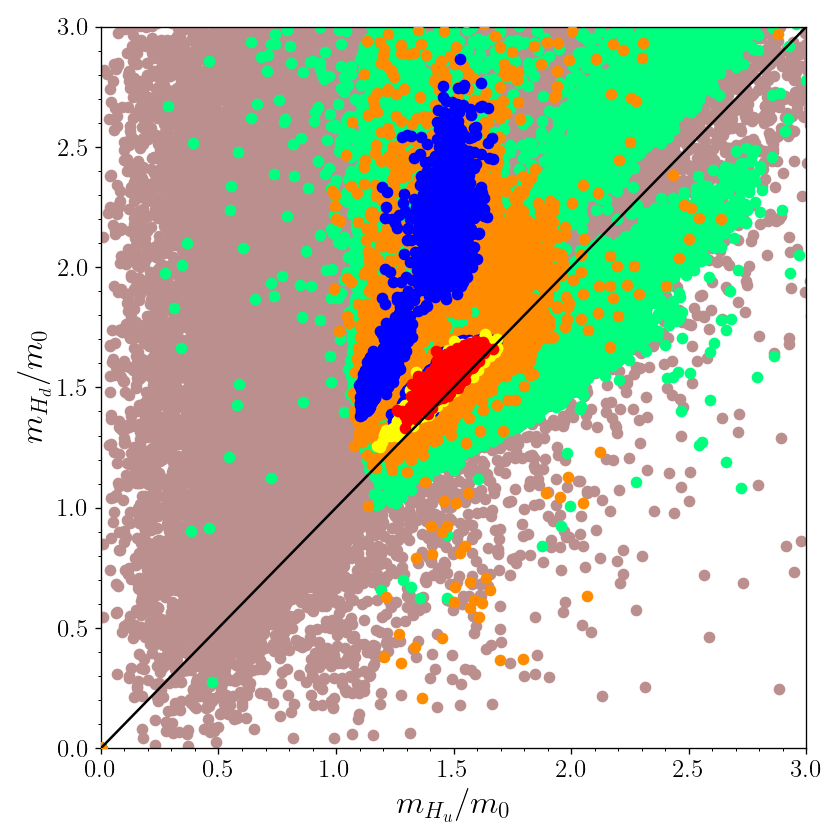}
\caption{Plots in the $\Rtbtau-\tan\beta$ and $m_{H_{d}}/m_{0}-m_{H_{u}}/m_{0}$ planes. All the points are compatible with REWSB and LSP neutralino condition. The green points are consistent with the mass bounds and constraints from rare $B-$meson decays. The acceptably fine-tuned solutions are shown in orange which form a subset of green. While the YU condition is not applied in the left plane, they are shown in blue in the right plane. Red and yellow points represent the solutions which are compatible with the DM constraints. The solutions compatible with the YU condition lie below the horizontal line in the left plane. The diagonal line in the right plane indicates the solutions where the GUT scale SSB mass terms of the MSSM Higgs fields are equal to each other.}
\label{fig:Rtbtau}
\end{figure}

As discussed in Section \ref{sec:intro}, the main impact on $\tan\beta$ parameter arises from the YU condition, and its strength can be seen in the $\Rtbtau-\tan\beta$ plane of Figure \ref{fig:Rtbtau}, which is plotted together with the ratios of the SSB masses of the Higgs fields to the sfermion masses. All the points are compatible with REWSB and LSP neutralino condition. The green points are consistent with the mass bounds and constraints from rare $B-$meson decays. The acceptably fine-tuned solutions are shown in orange which form a subset of green. While the YU condition is not applied in the left plane, they are shown in blue in the right plane. Red and yellow points represent the solutions which are compatible with the DM constraints. The solutions compatible with the YU condition lie below the horizontal line in the left plane. The diagonal line in the right plane indicates the solutions where the GUT scale SSB mass terms of the MSSM Higgs fields are equal to each other. While most of the solutions can be compatible with the constraints and acceptable fine-tuning condition, the YU solutions (below the horizontal line) can be realized only when $\tan\beta \gtrsim 40$. 

The results in the $m_{H_{d}}/m_{0}-m_{H_{u}}/m_{0}$ plane reveal an interesting relation between the GUT scale MSSM Higgs field masses. In the absence of NH terms, $y_{t}=y_{b}$ requires different values for $m_{H_{d}}$ and $m_{H_{u}}$ to be consistent with REWSB \cite{Baer:2009ie}. The diagonal line in this plane shows the regions where $m_{H_{d}}=m_{H_{u}}$, and there exist lots of solutions around this line. The NH terms open up another region in the parameter space compatible with the YU condition, which cannot be observed in the SUSY GUT models with the universal boundary conditions. In this context, the NH terms can also reduce the number of free parameters by setting $m_{H_{u}}=m_{H_{d}}$ and since there are still solutions compatible with the YU condition, the minimally constructed Yukawa unified GUT models can be classified now in NUHM Type 1 (NUHM with $m_{H_{u}}=m_{H_{d}}$ at $\mgut$). Moreover, the region with $m_{H_{d}}/m_{0} = m_{H_{u}}/m_{0} = 1$ represents the solutions with $m_{0}=m_{H_{d}}=m_{H_{u}}$ which corresponds to constrained MSSM (CMSSM) models; however, in our scans the minumum value for these ratios can be realized at about $m_{H_{d}}/m_{0} = m_{H_{u}}/m_{0} \simeq 1.2$, which can be fairly named as nearly CMSSM class. These relations are observed because the NH terms enhance the contributions from the sbottom to $H_{u}$, which triggers a significant separation between $H_{u}$ and $H_{d}$ masses through RGEs, and one can observe YU solutions consistent with REWSB even $m_{H_{u}}=m_{H_{d}}$ at $\mgut$. Such an effect can also be observed if one imposes non-universal boundary conditions at the GUT scale (see, for instance, \cite{Gogoladze:2011ce,Gogoladze:2011aa}).

\begin{figure}[t!]
\centering
\includegraphics[scale=0.4]{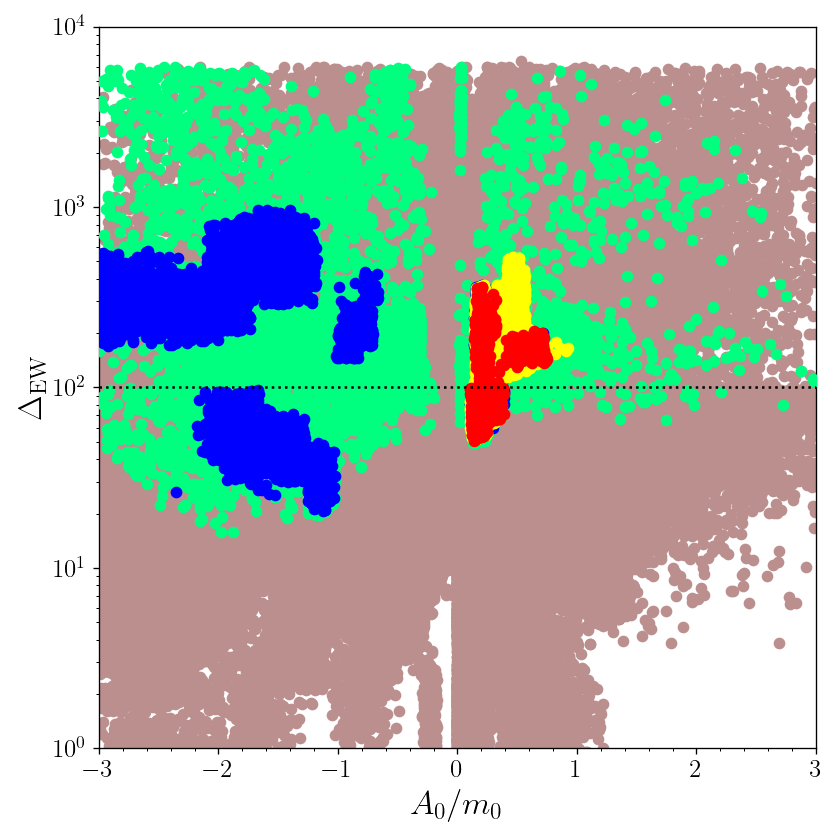}%
\includegraphics[scale=0.4]{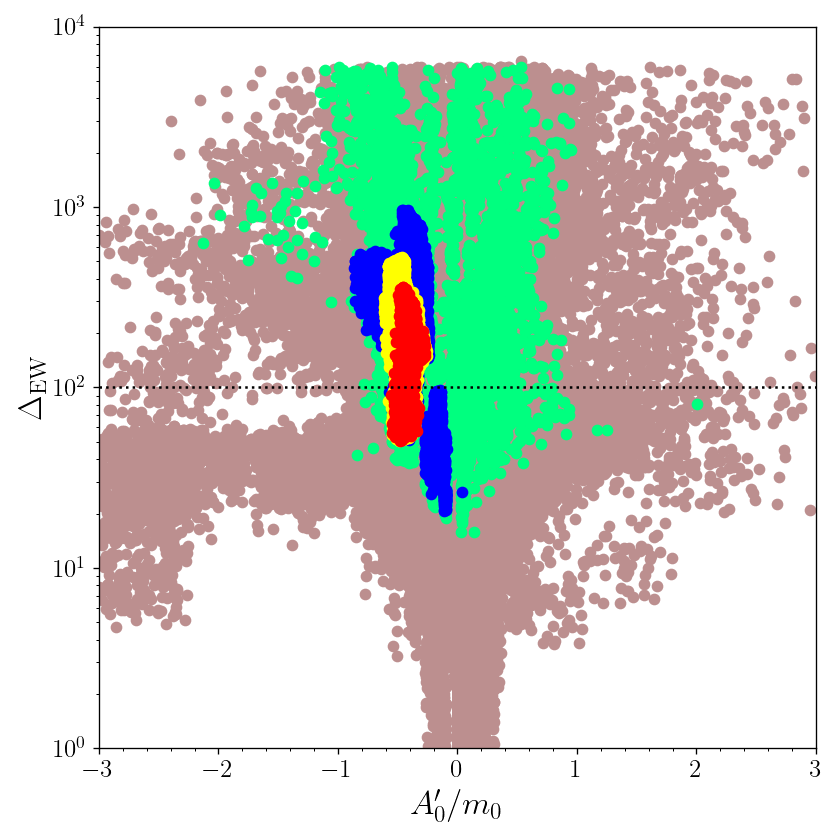}
\caption{The Fine-tuning results in correlation with the holomorphic (left) and NH (right) trilinear scalar couplings. The color coding is the same as Figure \ref{fig:FTfundamental}.}
\label{fig:FTA0s}
\end{figure}

The discussion continues in Figure \ref{fig:FTA0s} with the SSB trilinear scalar couplings from the holomorphic (left) and NH (right) parts of the SSB Lagrangian. The holomorphic $A_{0}$ term exhibits a typical behavior under the YU condition. YU favors the regions where $A_{0}$ is large compared with the SSB scalar mass $m_{0}$ (blue points). Indeed, it is required to be quite large when the universal boundary conditions are imposed \cite{Baer:2009ie}. In the scans, it is also observed that $A_{0}$ almost doubles the SSB scalar masses (blue points in $-2 \lesssim A_{0}/m_{0} \lesssim -1$). Comparing with the $\FTEW-A_{0}^{\prime}/m_{0}$ plane on the right, these large $A_{0}$ solutions correspond to the regions where NH trilinear couplings are nearly zero. However, with the inclusion of the NH terms, The YU solutions can also be realized consistent with the acceptable fine-tuning solutions for $0\lesssim A_{0}/m_{0}\lesssim 1$ where the trilinear couplings are relatively small ($0 \lesssim A_{0} \lesssim m_{0}$). Indeed, these solutions are also favored by the DM observations (red and yellow points). These DM favored solutions can be also seen in the $\FTEW-A_{0}^{\prime}/m_{0}$ plane in the range $-0.5\lesssim A_{0}^{\prime}/m_{0} \lesssim 0.1$.

\begin{figure}[t!]
\centering
\includegraphics[scale=0.4]{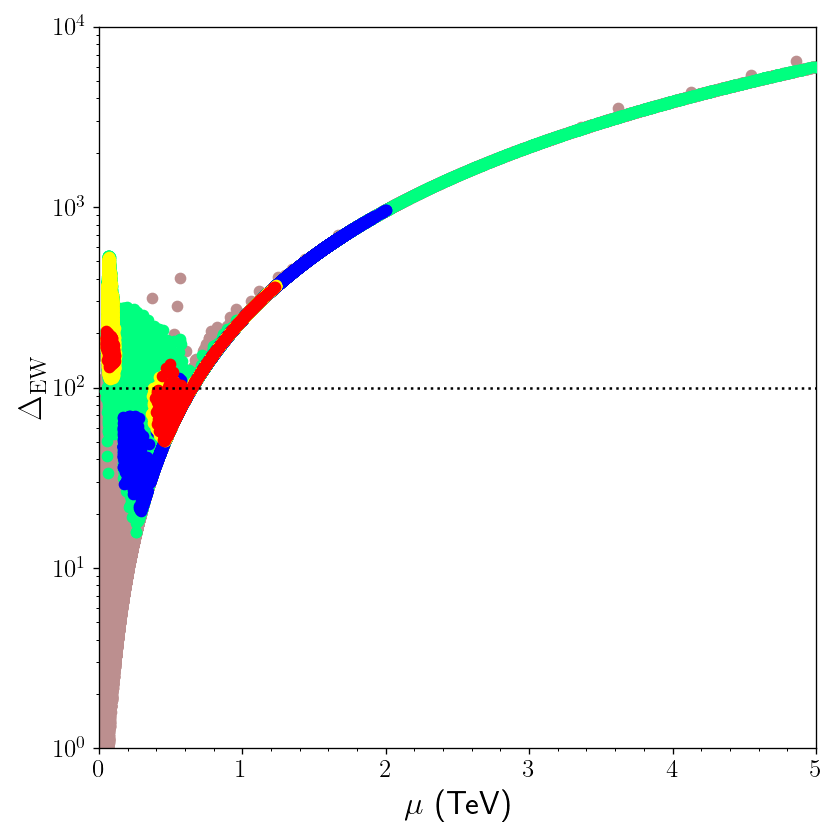}%
\includegraphics[scale=0.4]{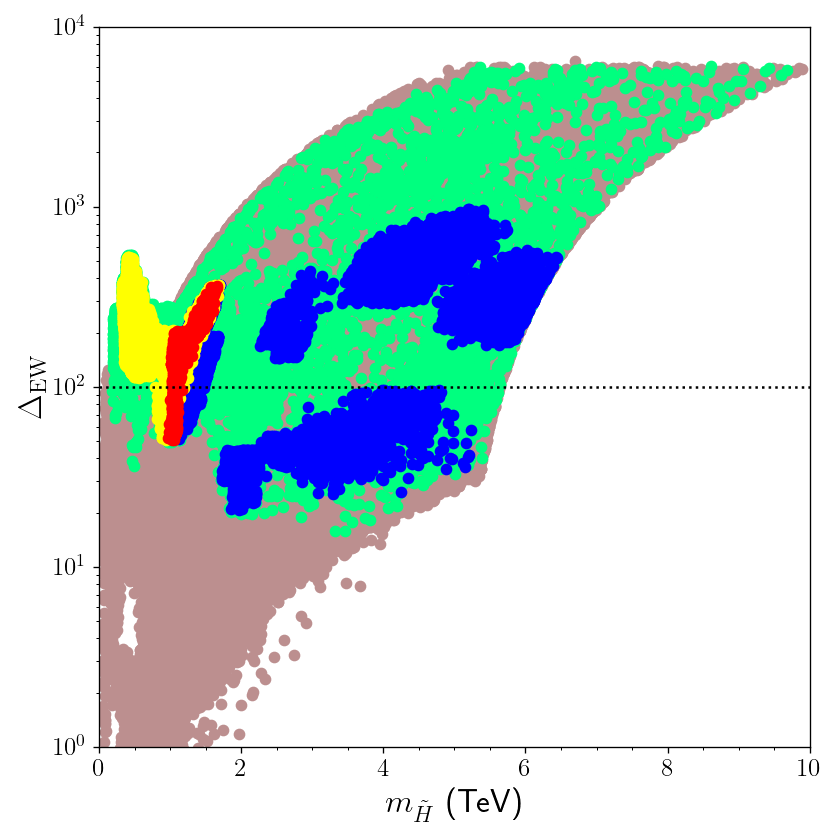}
\caption{The Fine-tuning measures in correlation with the $\mu-$term (left) and the Higgsino mass ($m_{\tilde{H}}$) (right). The color coding is the same as Figure \ref{fig:FTfundamental}.}
\label{fig:FTmus}
\end{figure}

As stated before, when the non-zero NH terms are imposed, the link between the Higgsino mass and the fine-tuning measure is removed which can be seen from the plots displayed in Figure \ref{fig:FTmus}, where $m_{\tilde{H}}$ represents the Higgsino mass in the right panel. The color coding is the same as Figure \ref{fig:FTfundamental}. The $\FTEW-\mu$ plane reveals the direct correlation between the $\mu-$term and $\FTEW$, which follows the solutions in the parabolic stream. Note that there is also a region in which the solutions deviates from the parabola when $\mu \lesssim 500$ GeV. In this region, since $\mu$ is relatively small, $m_{H_{u}}$ term in \ref{eq:MZ} takes over the $\mu-$term with the help of large $\tan\beta$, and $\FTEW$ is determined by the $C_{{H_{u}}}$ instead of $C_{\mu}$, which are given in Eq.(\ref{eq:CFT}). If the NH terms were not considered, the same correlation would be expected to be observed in the right panel of Figure \ref{fig:FTmus}. However, the presence of $\mu^{\prime}-$term breaks this correlation, as well as enhancing the Higgsino mass up to about 5 TeV, while the fine-tuning measure remains under 100. Also here the main impact comes from the DM constraints (red and yellow) which bound the Higgsino mass at about 2 TeV from above.

\section{Higgs and Sparticle Spectrum}
\label{sec:spectrum}

\begin{figure}[t!]
\centering
\includegraphics[scale=0.4]{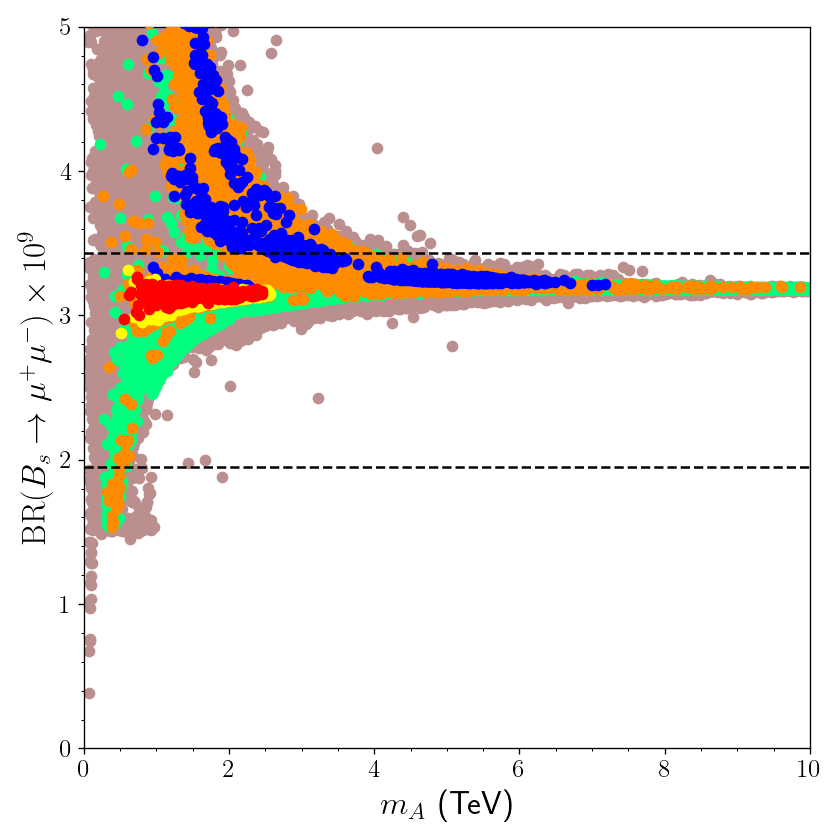}%
\includegraphics[scale=0.4]{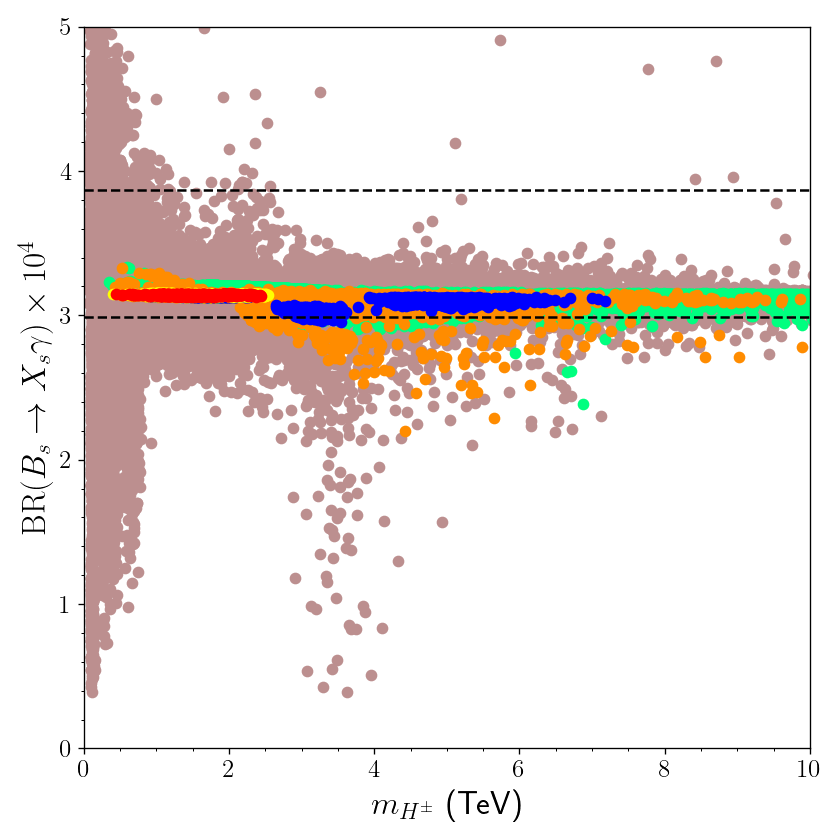}
\includegraphics[scale=0.4]{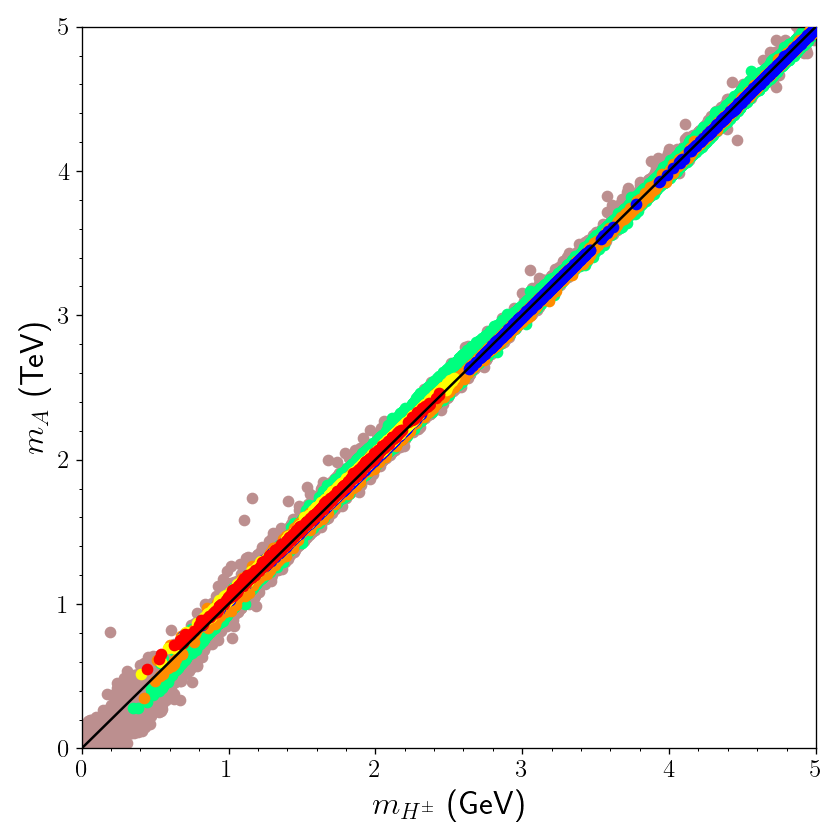}%
\includegraphics[scale=0.4]{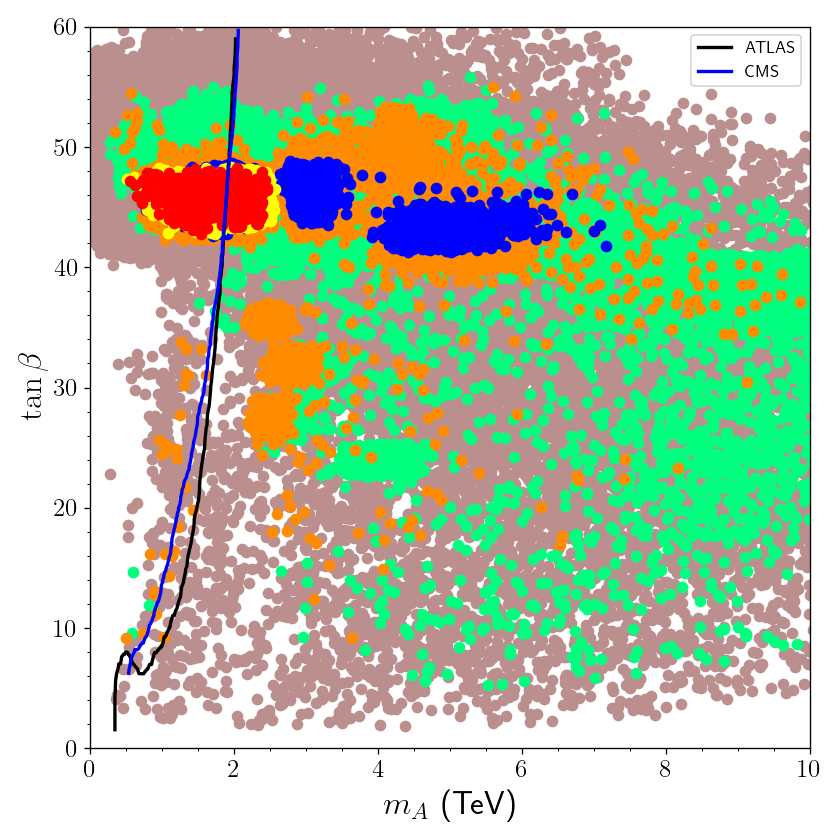}
\caption{Plots for the heavy Higgs bosons of MSSM in correlation with the rare $B-$meson decays (top), with each other (bottom-left) and $\tan\beta$ (bottom right). The color coding is the same as in Figure {\ref{fig:Rtbtau}}. The constraint from ${\rm BR}(B_{s}\rightarrow \mu^{+}\mu^{-})$ is not applied in the top-left panel, and that from ${\rm BR}(B\rightarrow X_{s}\gamma)$ is not applied in the top-right panel. Their experimental ranges as shown in Eq.(\ref{eq:constraints}) are represented by the horizontal dashed lines in these planes. In the $\tan\beta-m_{A}$ plane the solid curves represent the current bounds on the CP-odd Higgs boson from ATLAS \cite{ATLAS:2020zms} and CMS \cite{CMS:2022goy}.}
\label{fig:HiggsMasses}
\end{figure}

As mentioned above, the SM-like Higgs boson mass can receive contributions from NH terms by around 1 GeV or less, while the impact on the heavy Higgs boson can be realized much larger. The contributions from the NH terms to these heavy Higgs boson masses depend on the sign of $A_{0}^{\prime}$. Figure \ref{fig:HiggsMasses} displays the heavy Higgs boson masses together with some parameters and constraints such as rare $B-$meson decays (top) and $\tan\beta$ (bottom-right). The $m_{A}-m_{H^{\pm}}$ plane also shows the mass degeneracy between these Higgs bosons.  The color coding is the same as in Figure {\ref{fig:Rtbtau}}. The constraint from ${\rm BR}(B_{s}\rightarrow \mu^{+}\mu^{-})$ is not applied in the top-left panel, and that from ${\rm BR}(B\rightarrow X_{s}\gamma)$ is not applied in the top-right panel. Their experimental ranges as shown in Eq.(\ref{eq:constraints}) are represented by the horizontal dashed lines in these planes. In the $\tan\beta-m_{A}$ plane the solid curves represent the current bounds on the CP-odd Higgs boson from ATLAS \cite{ATLAS:2020zms} and CMS \cite{CMS:2022goy}. 

In the constrained SUSY GUT models (such as NUHM2 discussed in this work), these Higgs bosons are typically involved in the spectrum compatible with YU when they weigh more than about 1 TeV. However, with the contributions from the NH terms, their masses can be as light as about 400 GeV as seen from each plot of Figure \ref{fig:HiggsMasses}. However, solutions for such light Higgs bosons receive a strong impact from rare $B-$meson decays since the non-SM contributions to these decays are mediated by the extra scalars which couple to the quarks and leptons. The ${\rm BR}(B_{s}\rightarrow \mu^{+}\mu^{-})-m_{A}$ plane shows that the solutions with $m_{A}\lesssim 600$ GeV are excluded by the constraint on these rare decay mode of $B-$meson. Similarly, the neutral flavor changing process $B\rightarrow X_{s}\gamma$ receives extra contributions from the charged Higgs boson, but the ${\rm BR}(B\rightarrow X_{s}\gamma)-m_{H^{\pm}}$ shows that this decay mode of $B-$meson does not lead to a strong impact, since all the colored solutions place between the two horizontal dashed lines. Even though there is no direct correlation between ${\rm BR}(B_{s}\rightarrow \mu^{+}\mu^{-})$ and the charged Higgs mass, $A$ and $H^{\pm}$ are quite degenerate in mass in the consistent mass spectrum, thus ${\rm BR}(B_{s}\rightarrow \mu^{+}\mu^{-})$ constrains the charged Higgs mass by bounding $m_{A}$. 

Even though the NH terms can relax the spectrum without disturbing the consistency with the constraints, they can usually help to control the SUSY contributions to some processes and relax the impact from some exprerimental constraints which can probe the SUSY models indirectly. However, this situation cannot be observed if one considers a direct impact from LHC analyses. For instance, the current LHC analyses for $A,H\rightarrow \tau\tau$ events can exclude the solutions with $m_{A}\lesssim 2$ TeV for the large $\tan\beta$ \cite{ATLAS:2020zms,CMS:2022goy}. The NH terms cannot help much to the relatively light CP-odd and Charged Higgs boson solutions to be consistent with the current LHC results. The $\tan\beta - m_{A}$ plane shows the distribution of our solutions in this plane together with the exclusion curves from the current analyses. Even though the low fine-tuned solutions (orange) can be realized when these Higgs bosons as heavy as about 10 TeV, the YU condition (blue points) bounds their masses at about 7 TeV from above. The relic density constraint (red and yellow points) reduces this bound further to about 2.5 TeV, and most of these solutions lie in the left side of the exclusion curve in the $\tan\beta-m_{A}$ plane, which is excluded. Despite realizing a strong impact in the parameter space from these analyses, the strict bounds on the mass of CP-odd Higgs boson also make this model more predictive. The recent analyses have shown that the CP-odd Higgs boson solutions can be probed/tested up to about $m_{A}=2.5$ TeV through the $A,H\rightarrow \tau\tau$ events when LHC starts operating with high luminosity (with $\mathcal{L}\simeq 3000$ fb$^{-1}$) \cite{CMS:2018oxh}.

\begin{figure}[t!]
\centering
\includegraphics[scale=0.4]{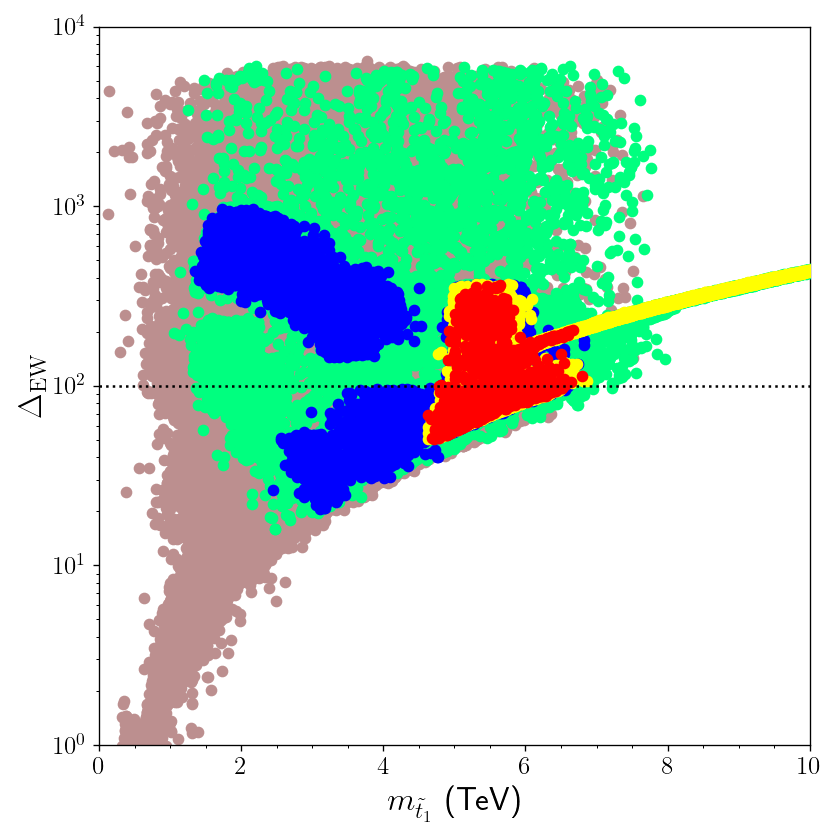}%
\includegraphics[scale=0.4]{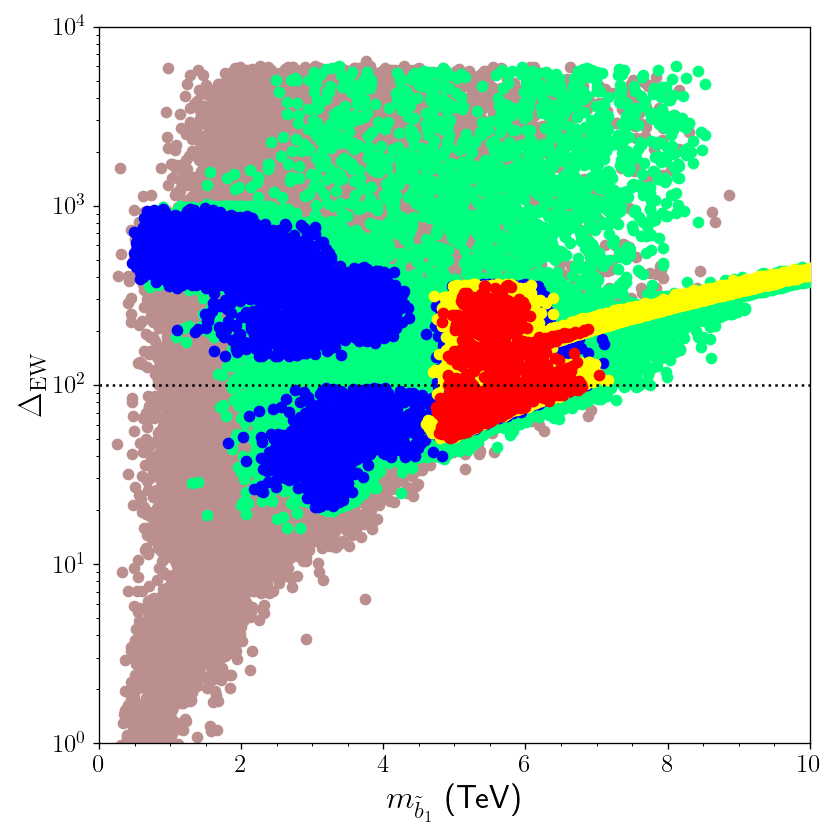}
\includegraphics[scale=0.4]{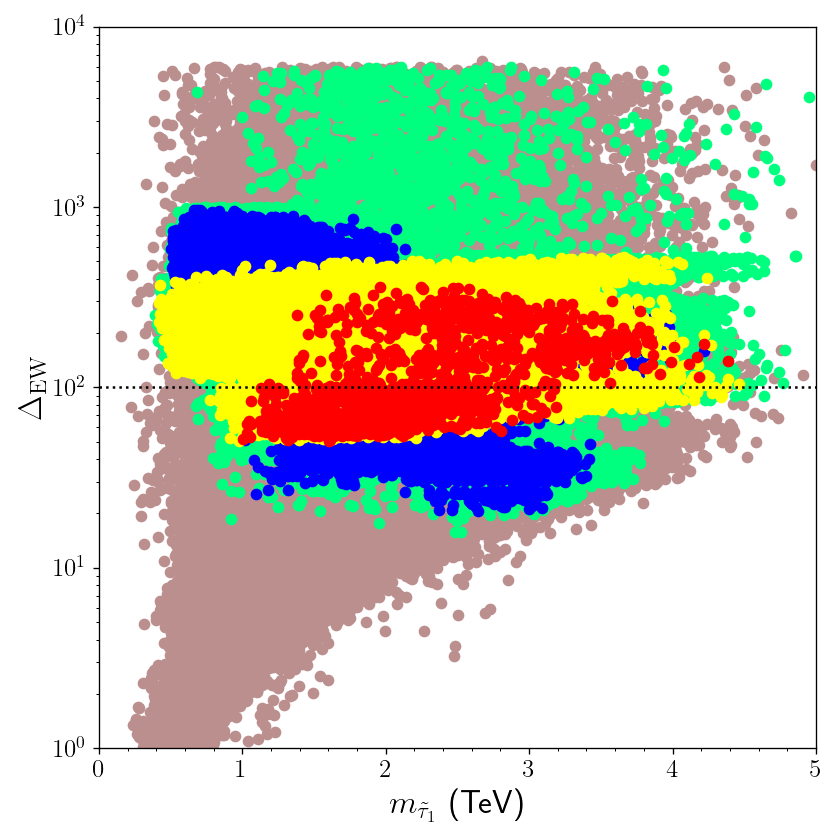}%
\includegraphics[scale=0.4]{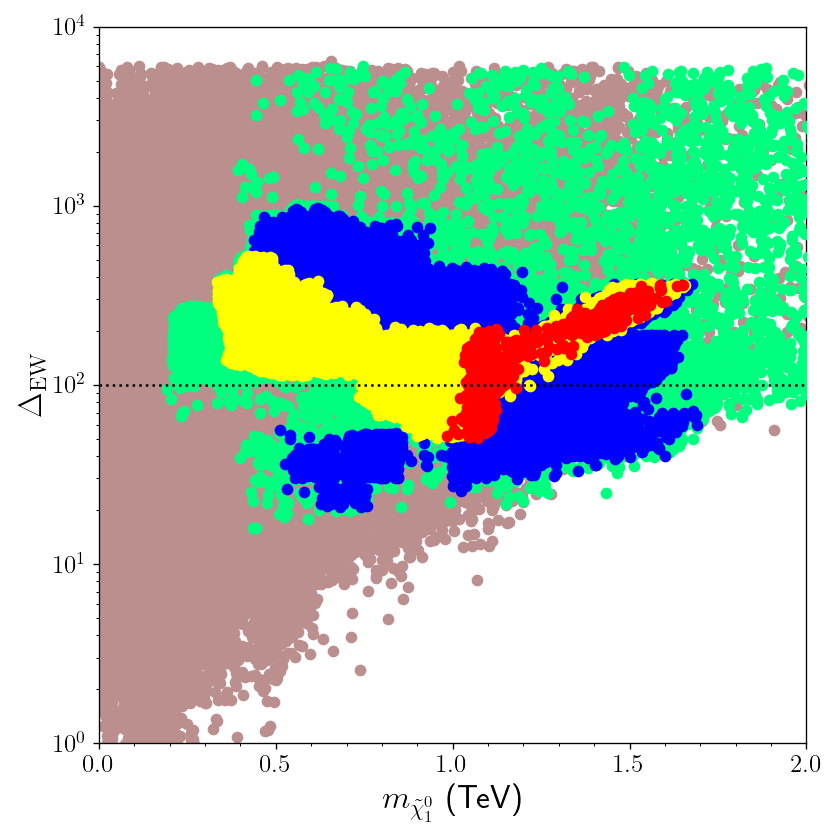}
\caption{The Fine-tuning results and the supersymmetric mass spectra involving stop (top-left), sbottom (top-right), stau (bottom-left) and the LSP neutralino (top-right). The color coding is the same as in Figure \ref{fig:FTfundamental}.}
\label{fig:sparticles}
\end{figure}

Figure \ref{fig:sparticles} displays the results for the mass spectrum of the supersymmetric particles involving stop (top-left), sbottom (top-right), stau (bottom-left) and the LSP neutralino (bottom-right). The color coding is the same as in Figure \ref{fig:FTfundamental}. The stop and sbottom are also of a special importance in realizing the consistent SM-like Higgs boson mass in the spectrum. In the absence of the NH terms, these particles are needed to be heavier to accommodate a 125 GeV Higgs boson, and the necessary radiative contributions to the Higgs boson can be realized when $m_{\tilde{t}} \gtrsim 1$ TeV and $m_{\tilde{b}} \gtrsim 4-5$ TeV. The interference of the NH terms in the stop sector is suppressed by the VEV of $H_{d}$, thus there is no much expectation to change the needed mass scales of the stops, and indeed, as seen from the $\FTEW-m_{\tilde{t}}$ plane, the minimum stop mass allowed by the LHC constraints (green) can accommodate the stop mass as $m_{\tilde{t}} \gtrsim 1$ TeV. The YU condition (blue) can shift this bound further to about 2 TeV, and the relic density solutions start to appear in the region where $m_{\tilde{t}}\gtrsim 4.4$ TeV. The top-right panel displays the solutions for the sbottom mass, and it can be seen that similar mass bounds can hold also for the sbottom. However, as mentioned above, a typical mass bound is much further than those realized in the analyses represented in this work. In our results, the sbottom and stop can weigh at the comparable mass scales when the NH terms interfere in the mixing of these particles.  

In addition to the squarks, the bottom planes of Figure \ref{fig:sparticles} display the masses of the particles which are more relevant to the electroweak sector such as stau (left) and the LSP neutralino (right). In contrast to the squarks, the stau can be as light as about 1 TeV compatible with the constraints including the YU and relic density in the low fine-tuned region. The LSP neutralino mass shown in the bottom-right panel can lie in a wide range from about 300 GeV to 1.8 TeV. Some regions in this range can yield low (yellow) or consistent (red) relic density solutions, while most of the low -fine-tuned solutions yield large relic density (blue) which are problematic in DM analyses. The distributions of the colors in the $\FTEW-m_{\tilde{\chi}_{1}^{0}}$ seems an interesting pattern in which the colors can be distinguished sharply from each other. Such a pattern happens because of the different species of LSP neutralino. The low relic density (yellow) and consistent DM (red) solutions are realized when the LSP neutralino is formed mostly by Higgsinos. In this case, the coannihilation and annihilation processes take over to reduce the relic density of LSP neutralino and depending on their strength, one can realize either consistent or lower relic density solutions. On the other hand, LSP neutralino can also be formed by Bino. Bino-like LSP typically leads to a very large relic density, and one needs to identify several coannihilation processes to reduce its relic density down to the ranges set by the Planck measurements. However, relatively more restricted SUSY models do not allow most of the coannihilation scenarios and the existent coannihilation scenarios are usually not enough to reduce its relic density to the compatible ranges. Considering the color distribution in the $\FTEW-m_{\tilde{\chi}_{1}^{0}}$ one can easily deduce the species of LSP, but it will be discussed in more details in the next section.

\section{DM Implications}
\label{sec:DM}

As described in Section \ref{sec:scan}, the scans accept only the solutions in which one of the neutralinos are involved in the spectrum as to be LSP to accommodate suitable DM candidates. Even though MSSM can provide compelling candidates from the weakly interacting massive particles (WIMPs), the assumption that the DM relic density is saturated only by the LSP neutralino leads to a strong impact in the parameter space. As discussed in the previous section, despite wide regions compatible with the LHC constraints, the relic density requirement (red) can be met only in a small region. In this context, also the low relic density solutions (yellow) are displayed which can be consistent DM candidate in non-standard scenarios such as those with multiple DM candidates.

\begin{figure}[t!]
\centering
\includegraphics[scale=0.4]{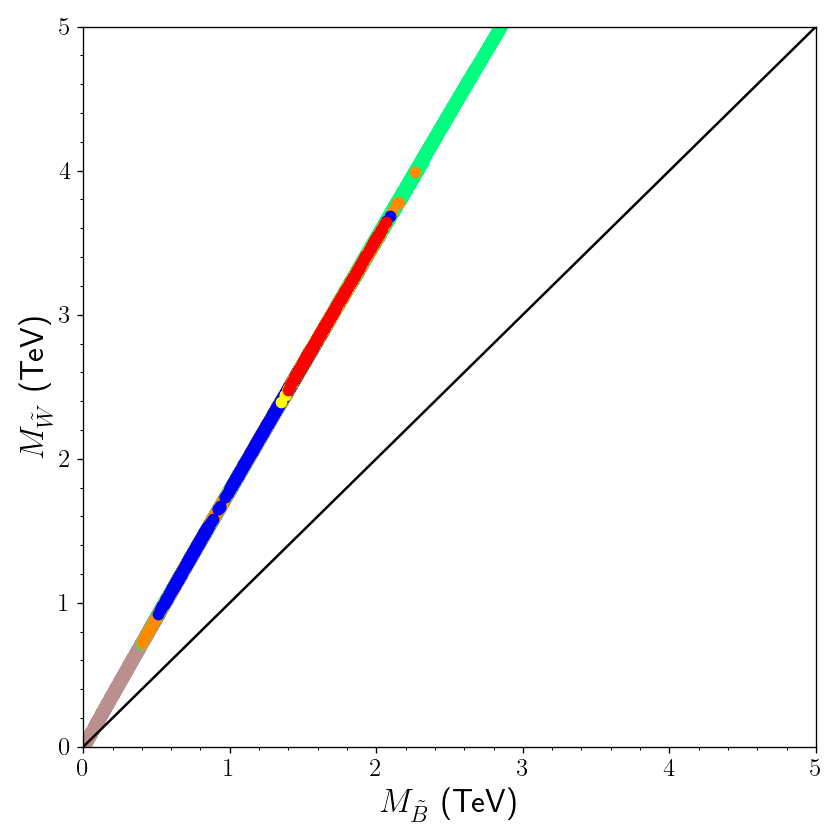}%
\includegraphics[scale=0.4]{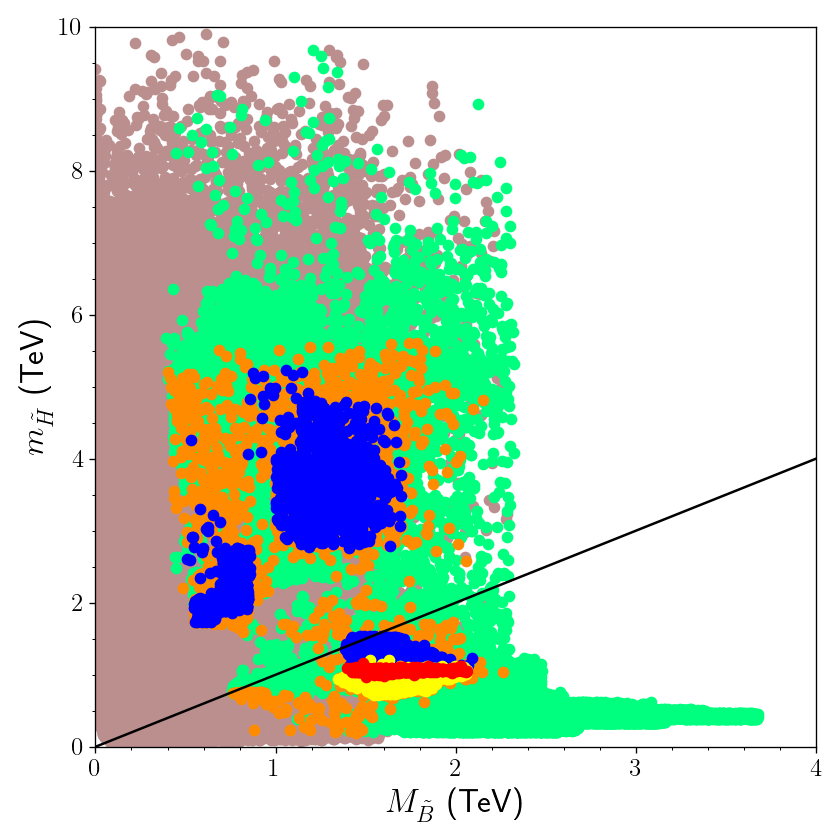}
\caption{Plots for the Wino (left) and Higgsino (right) masses in correlation with the Bino mass. The color coding is the same as in Figure \ref{fig:Rtbtau}. The diagonal lines show the regions where the plotted masses are degenerate with each other.}
\label{fig:LSPspecies}
\end{figure}

This requirement makes the MSSM gauginos and Higgsinos more relevant in our analyses since they form the LSP neutralino. Figure \ref{fig:LSPspecies} display their masses individually with plots for the Wino (left) and Higgsino (right) masses in correlation with the Bino mass. The color coding is the same as in Figure \ref{fig:Rtbtau}. The diagonal lines show the regions where the plotted masses are degenerate with each other. Since our model is restricted to impose universal gaugino masses at $\mgut$, the Wino and Bino masses exhibit almost a linear correlation between each other as $M_{\tilde{W}}\simeq 2M_{\tilde{B}}$ as seen from the $M_{\tilde{W}}-M_{\tilde{B}}$ plane. In this context, the Wino cannot take part in the composition of LSP neutralino in our solutions. On the other hand, the Higgsino and Bino masses are not linked to each other in the boundary conditions imposed in our scans, thus there is no a specific correlation between their mases. The $m_{\tilde{H}}-m_{\tilde{B}}$ plane shows that the LSP happens to be Bino-like in the solutions above the diagonal line, while the Higssino takes over in those below. Around the diagonal line $m_{\tilde{H}}\simeq M_{\tilde{B}}$ and these solutions predict a Bino-Higgsino mixture in the LSP composition. 

\begin{figure}[t!]
\centering
\includegraphics[scale=0.4]{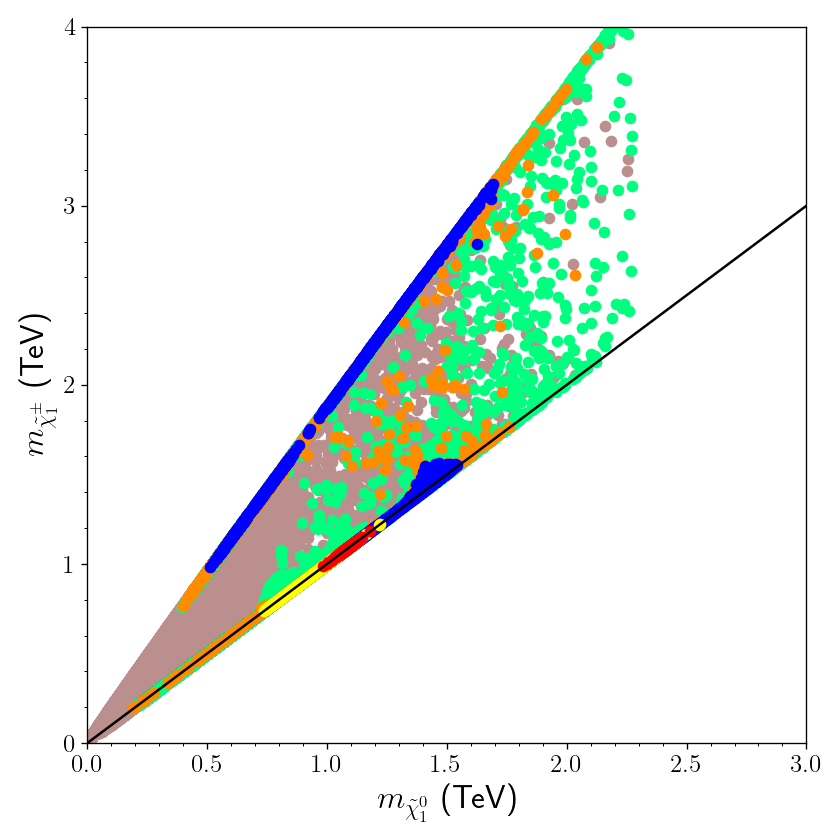}%
\includegraphics[scale=0.4]{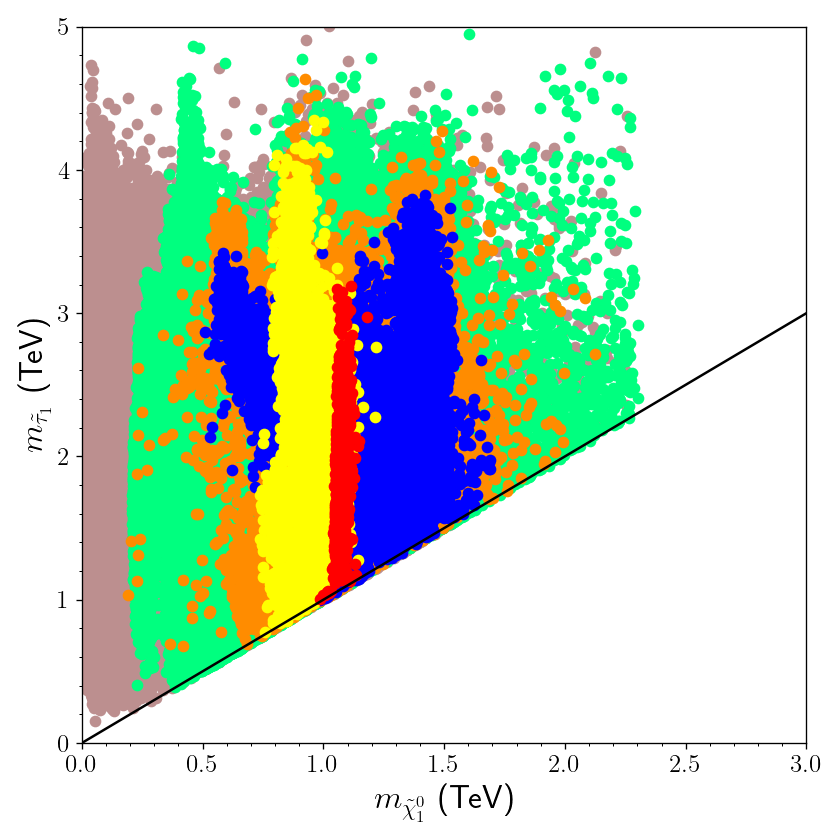}
\includegraphics[scale=0.4]{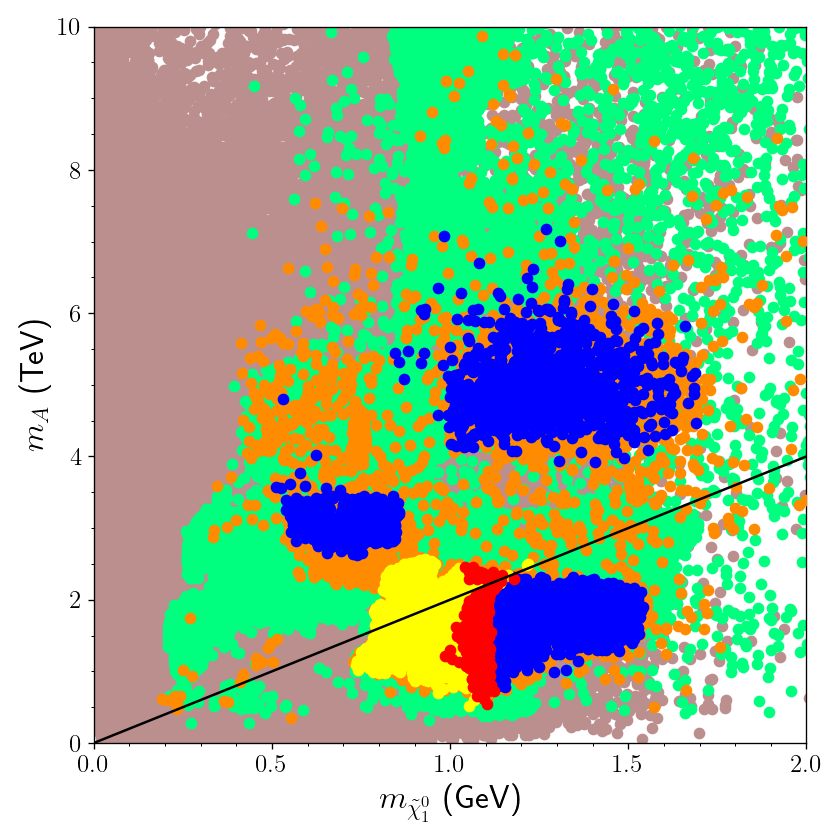}%
\includegraphics[scale=0.4]{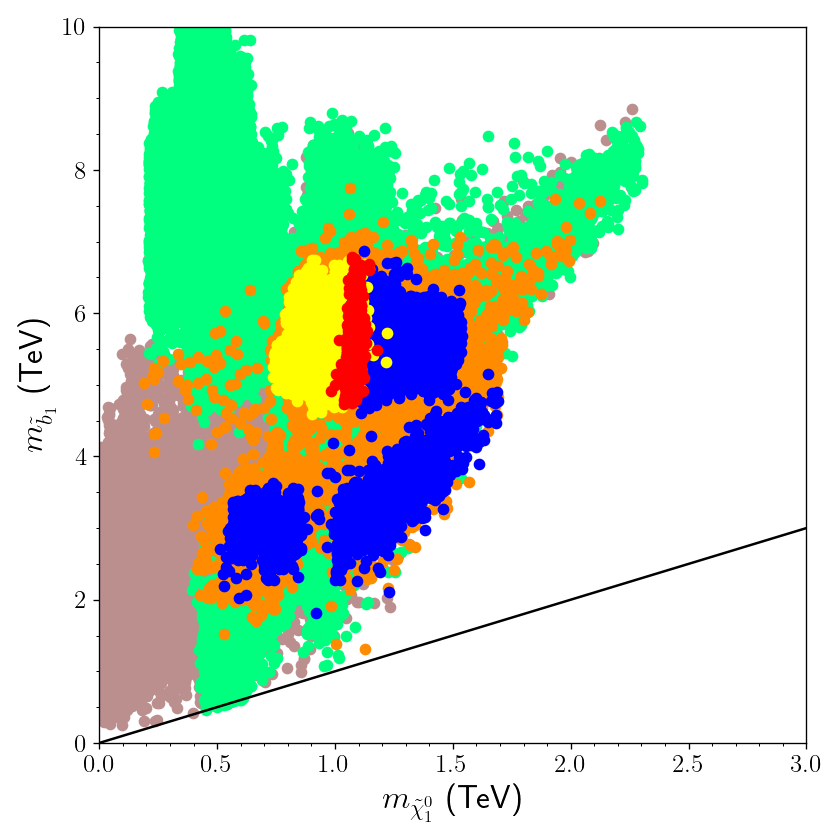}
\caption{Plots in the $m_{\tilde{\chi}_{1}^{\pm}}-m_{\tilde{\chi}_{1}^{0}}$, $m_{\tilde{\tau}_{1}}-m_{\tilde{\chi}_{1}^{0}}$, $m_{A}-m_{\tilde{\chi}_{1}^{0}}$ and $m_{\tilde{b}_{1}}-m_{\tilde{\chi}_{1}^{0}}$ planes. The color coding is the same as in Figure \ref{fig:Rtbtau}. The diagonal lines indicate the regions where the plotted particles are degenerate in mass except in the $m_{A}-m_{\tilde{\chi}_{1}^{0}}$ plane. The diagonal line in this plane shows the regions of $A-$resonance where $m_{A}=2m_{\tilde{\chi}_{1}^{0}}$.}
\label{fig:coan}
\end{figure}

The sharp distinction among the colors (yellow, red and blue) can also be recognized in the $m_{\tilde{H}}-m_{\tilde{B}}$ plane. The low (yellow) and desired (red) relic density are observed rather in the regions where the LSP is Higgsino-like. When Bino takes part in the LSP composition, on the other hand (around the diagonal line and above), the relic density of LSP neutralino becomes larger. This can be understood by exploring the possible coannihilation and annihilation scenarios which are represented in Figure \ref{fig:coan} with plots in the $m_{\tilde{\chi}_{1}^{\pm}}-m_{\tilde{\chi}_{1}^{0}}$, $m_{\tilde{\tau}_{1}}-m_{\tilde{\chi}_{1}^{0}}$, $m_{A}-m_{\tilde{\chi}_{1}^{0}}$ and $m_{\tilde{b}_{1}}-m_{\tilde{\chi}_{1}^{0}}$ planes. The color coding is the same as in Figure \ref{fig:Rtbtau}. The diagonal lines indicate the regions where the plotted particles are degenerate in mass except in the $m_{A}-m_{\tilde{\chi}_{1}^{0}}$ plane. The diagonal line in this plane shows the regions of $A-$resonance where $m_{A}=2m_{\tilde{\chi}_{1}^{0}}$. The $m_{\tilde{\chi}_{1}^{\pm}}-m_{\tilde{\chi}_{1}^{0}}$ plane reveals two streams of the solutions. In one stream the solutions are accumulated around the diagonal line where the chargino and the LSP neutralino masses are nearly degenerate. Such solutions are typical for the Higgsino-like LSP, since in those cases the lightest chargino is also formed by the Higgsinos. All the consistent DM solutions are placed in this stream. In the second stream the chargino happens to be much heavier than the LSP neutralino. Such solutions represent the Bino-like LSP, and since there is no degeneracy between the chargino and LSP neutralino mass, the relic density of the Bino-like LSP cannot be reduced through the chargino-neutralino coannihilation scenario.

Another possible scenario can be identified as the stau-neutralino coannihilation processes. The $m_{\tilde{\tau}_{1}}-m_{\tilde{\chi}_{1}^{0}}$ shows that stau-neutralino coannihilation solutions can be observed for both the Higgsino-like and Bino-like LSP. However, one can here see again the distinction among the colors. When the LSP is Higgsino-like, the coannhilation processes among these particles happen through the Yukawa interactions, while Bino-like LSP participates these coannihilation processes with the $U(1)_{Y}$ gauge coupling $g_{1}$. The hierarchy between the couplings also determines whether a solution can or cannot be a suitable DM. Since $y_{\tau} > g_{1}$, the stau-neutralino coannihilation processes yield a stronger reduction in the LSP relic density and it yields to a consistent (red) or low (yellow) relic density. On the other hand, these processes are weak in the Bino-like LSP solutions, thus they cannot accommodate compatible relic density for the LSP neutralino, so the solutions typically lead to a large relic density for DM. Also the CP-odd Higgs boson in the $m_{A}-m_{\tilde{\chi}_{1}}^{0}$ plane is shown in which the diagonal line corresponds to solutions with $m_{A}=2m_{\tilde{\chi}_{1}^{0}}$. In this region the CP-odd Higgs boson makes a resonance with a pair of the LSP neutralinos which triggers annihilation of the LSPs into a CP-odd Higgs boson. Such $A-$resonance solutions are also very effective in reducing the relic density, but as seen from the distributions of colors in this plane, this scenario can lead to compatible relic density solutions only in the regions where LSP is Higgsino-like. Sbottom mass with respect to the LSP neutralino mass is given in the top-right plane with a diagonal line representing the mass degeneracy between them. Even though one can identify sbottom-neutralino coannihilation solutions they can happen only when LSP is bino-like. The accepted relic density ranges (yellow and red) can allow sbottom to be as light as about 4-5 TeV only.

\begin{figure}[t!]
\centering
\includegraphics[scale=0.4]{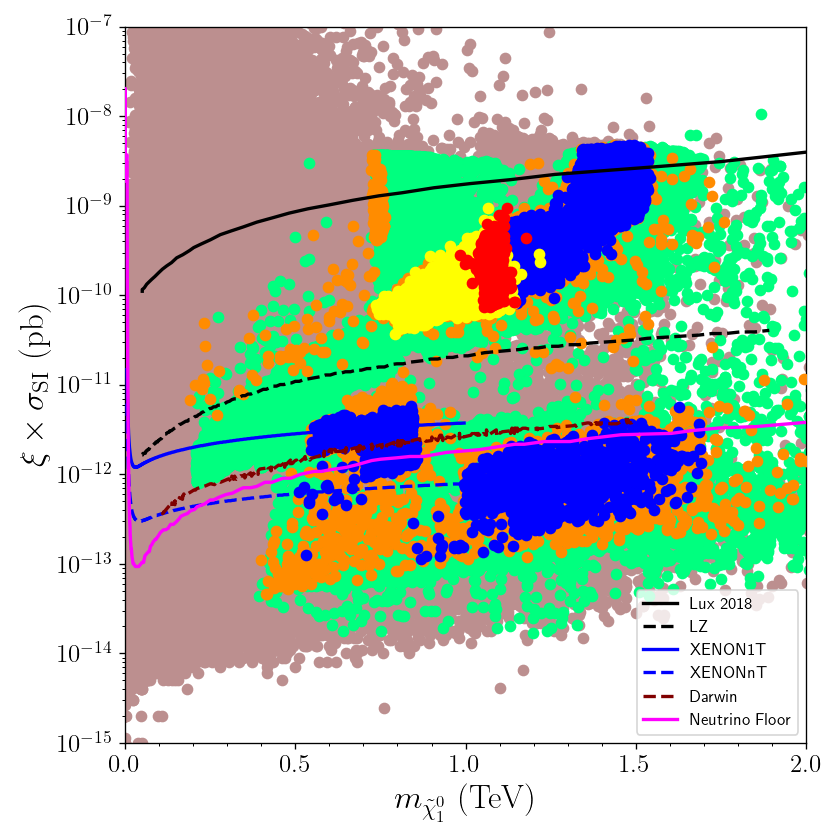}%
\includegraphics[scale=0.4]{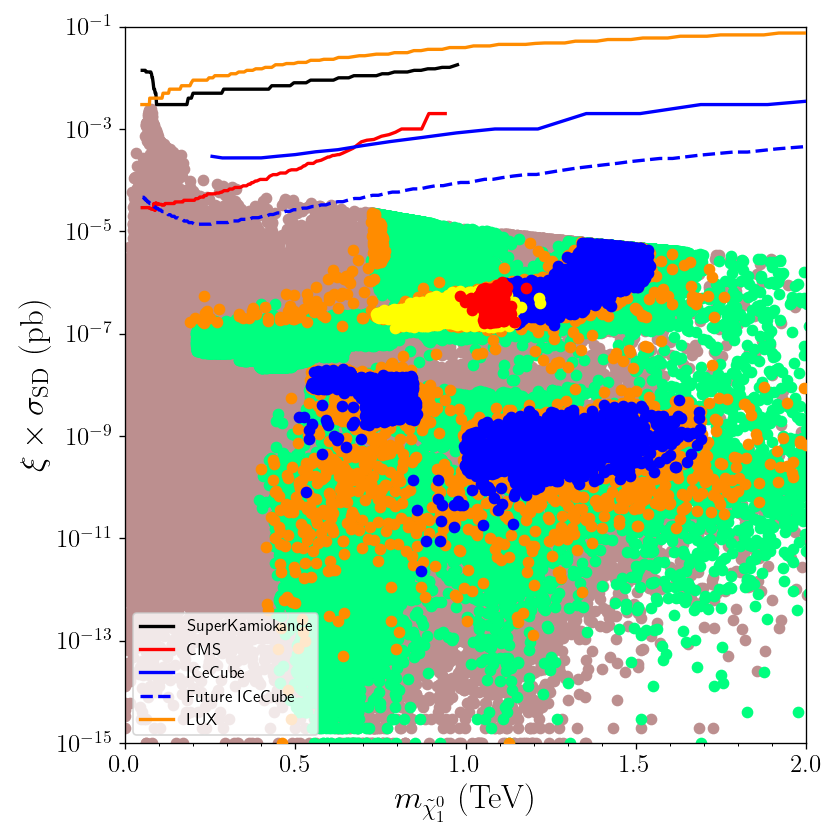}
\caption{Spin-independent (left) and spin-dependent (right) scattering cross-section results for the DM scattering events in correlation with the LSP neutralino mass. The curves represent the current bounds (solid) and future projections (dashed) of some experiments \cite{Akerib:2018lyp,Aprile:2020vtw,Aalbers:2016jon,Tanaka:2011uf,Khachatryan:2014rra,Abbasi:2009uz,Akerib:2016lao}. These experiments with their corresponding curves are listed in the legends in each plot.}
\label{fig:SISD}
\end{figure}

The Higgsino-like and Bino-like LSP can also be distinguished in direct detection of DM experiments operating over the possible scatterings of DM at nuclei. In these experiments, the Higgsino-like LSP is typically expected to lead to large cross-sections in the scattering events, and thus they take the strongest negative impact from these experiments. On the other hand, Bino-like LSP is usually scattered by nuclei with low cross-section, and one may need to wait little longer for the experiments to improve their sensitivity. The scattering cross-section results in the spin-independent (left) and spin-dependent scattering events are displayed in Figure \ref{fig:SISD} in correlation with the LSP mass. The curves represent the current bounds (solid) and future projections (dashed) of some experiments \cite{Akerib:2018lyp,Aprile:2020vtw,Aalbers:2016jon,Tanaka:2011uf,Khachatryan:2014rra,Abbasi:2009uz,Akerib:2016lao}. These experiments with their corresponding curves are listed in the legends in each plot. It should be noted here that these exclusion curves from the experiments are derived by assuming the DM candidate has the correct relic density pointed by the Planck measurements. Even though this assumption holds for the red points, the scattering cross-sections should be rescaled for the low relic density solutions (yellow). Since the low relic density solutions cannot be accounted for the whole observations in the DM experiments, their scattering cross-sections are rescaled by $\xi$, which is defined as follows \cite{Belanger:2015vwa}:

\begin{equation}
\setstretch{2.5}
\xi = \left\lbrace \begin{array}{ll}
1 & {\rm for~} 0.114 \leq \Omega h^{2} \leq 0.126 \\
\dfrac{\Omega h^{2}}{0.12} & {\rm for}~ \Omega h^{2} < 0.114
\end{array}\right.
\label{eq:xi}
\end{equation}

As can be seen from the $\sigma_{{\rm SI}}-m_{\tilde{\chi}_{1}^{0}}$, all the relic density solutions yield to considerably large cross-sections in the scattering events. Even though the results from the XENON1T experiment are more sensitive to such solutions, its sensitivity covers only the LSP mass up to about 1 TeV. Thus, the relic density solutions, which are Higgsino-like LSP, are expected to be tested soon in LUX-Zeplin (LZ) experiment. One can easily identify the Bino-like LSP solutions, since they lie in the regions with low scattering cross-section. Since these solutions yield large relic density, they need to be modified and/or embedded into non-standard DM scenarios to be compatible with the latest Planck measurements. In such cases the further upgrades in XENON experiment might be able to probe such solutions. However, most of the Bino-like solutions lie below the neutrino floor curve. In this region, the main background is formed by the neutrino scatterings, and up to some energy levels neutrinos can imitate the DM. Thus, in addition to improving the sensitivity, one needs more statistics in this region to be able to distinguish the DM candidate from the neutrinos.

\begin{table}[h!]
\setstretch{1.2}
\scalebox{0.7}{
\begin{tabular}{|c|c|c|c|c|c|c|}
\hline  & Point 1 & Point 2 & Point 3 & Point 4 & Point 5 & Point 6 \\ \hline
$m_{0}$ & 4492 & 4733 & 5055 & 4726 & 5581 & 5496 \\
$M_{1/2}$ & 1514 & 1158 & 4648 & 4440 & 4665 & 3120 \\
$A_{0}/m_{0}$ & -1.13 & -2.35 & 0.212 & 0.235 & 0.231 & 0.184 \\
$A_{0}^{\prime}/m_{0}$ & -0.108 & 0.037 & -0.537 & -0.484 & -0.514 & -0.326 \\
$\tan\beta$ & 47.54 & 46.58 & 44.66 & 47.24 & 45.54 & 48.17 \\
$m_{H_{d}}$ & 6895 & 8887 & 8079 & 8059 & 8917 & 7879 \\
$m_{H_{u}}$ & 5179 & 5961 & 7996 & 7441 & 8493 & 6812 \\ \hline
$\FTEW$ & 20.69 & 26.18 & 97.14 & 96.04 & 99.95 & 95.03 \\
$\mu$ & 293.3 & 256.1 & 634.4 & 631.9 & 644.6 & 628.6 \\
$\mu^{\prime}$ & 1680 & 3997 & 416.6 & 445.5 & 424.1 & 692.7 \\ \hline
$m_{h}$ & 123.1 & \cred{125.6} & 124.2 & 124 & 124.2 & 123 \\
$m_{H}$ & 2841 & 4799 & 2024 & 2070 & 2275 & 2069 \\
$m_{A}$ & 2841 & 4798 & 2024 & 2070 & \cred{2275} & 2069 \\
$m_{H^{\pm}}$ & 2849 & 4785 & 1973 & 2036 & 2226 & 2062 \\ \hline
$m_{\tilde{\chi}_{1}^{0}}$,$m_{\tilde{\chi}_{2}^{0}}$ & \cred{682.3}, 1294 & \cred{531.2}, 1022 & \cred{1072}, \cred{1075} & \cred{1096}, \cred{1100} & \cred{1090}, \cred{1093} & \cred{1327}, \cred{1343} \\
$m_{\tilde{\chi}_{3}^{0}}$,$m_{\tilde{\chi}_{4}^{0}}$ & 1980, 1984 & 4194, 4195 & 2118, 3891 & 2020, 3718 & 2129, 3912 & 1424, 2628 \\
$m_{\tilde{\chi}_{1}^{\pm}}$,$m_{\tilde{\chi}_{2}^{\pm}}$ & 1294, 1985 & 1022, 4195 & \cred{1074}, 3891 & \cred{1099}, 3718 & \cred{1092}, 3912 & \cred{1340}, 2628 \\ \hline
$m_{\tilde{g}}$ & 3441 & 2741 & 9354 & 8953 & 9413 & 6559 \\
$m_{\tilde{u}_{1}}$,$m_{\tilde{u}_{2}}$ & 5159, 5302 & 5021, 5222 & 9247, 9599 & 8787, 9143 & 9542, 9901 & 7599, 7833 \\
$m_{\tilde{t}_{1}}$,$m_{\tilde{t}_{2}}$ & 3110, 3316 & 2444, 2737 & 6576, 7254 & 6356, 6914 & 6678, 7272 & 5306, 5612 \\ \hline
$m_{\tilde{d}_{1}}$,$m_{\tilde{d}_{2}}$ & 5248, 5302 & 5218, 5223 & 9220, 9599 & 8782, 9143 & 9537, 9902 & 7635, 7833 \\
$m_{\tilde{b}_{1}}$,$m_{\tilde{b}_{2}}$ & 3045, 3269 & 2185, 2625 & 6898, 7250 & 6454, 6908 & 6809, 7267 & 5251, 5602 \\ \hline
$m_{\tilde{\nu}_{e}}$,$m_{\tilde{\nu}_{\tau}}$ & 4532, 3653 & 4672, 3453 & 5845, 4644 & 5479, 4238 & 6291, 4914 & 5799, 4681 \\
$m_{\tilde{e}_{1}}$,$m_{\tilde{e}_{2}}$ & 4533, 4631 & 4673, 4958 & 5356, 5846 & 5057, 5480 & 5890, 6292 & 5688, 5799 \\
$m_{\tilde{\tau}_{1}}$,$m_{\tilde{\tau}_{2}}$ & 2644, 3655 & 2135, 3454 & 1870, 4646 & \cred{1179}, 4239 & 1956, 4915 & 2980, 4683 \\ \hline
$\Omega h^{2}$ & \cblue{105.2} & \cblue{133.1} & \cred{0.117} & \cred{0.126} & \cred{0.114} & \cblue{0.187} \\
$\sigma_{{\rm SI}}$ & $ 3.19 \times 10^{-12} $ & $ 1.25 \times 10^{-13} $ & $ 6.53 \times 10^{-11} $ & $ 7.85 \times 10^{-11} $ & $ 6.42 \times 10^{-11} $ & $ 2.3 \times 10^{-9} $ \\
$\sigma_{{\rm SD}}$ & $ 1.06 \times 10^{-8} $ & $ 8.18 \times 10^{-11} $ & $ 1.42 \times 10^{-7} $ & $ 1.65 \times 10^{-7} $ & $ 1.38 \times 10^{-7} $ & $ 3.33 \times 10^{-6} $ \\ \hline
LSP-$\tilde{B}\%$ & \cblue{99.92} & \cblue{99.99} & 0.093 & 0.119 & 0.094 & \cblue{12.23} \\
LSP-$\tilde{H}\%$ & 0.075 & 0.012 & 99.87 & 99.83 & 99.87 & 87.6 \\ \hline
$y_{t}$,$y_{b}$,$y_{\tau}$ & 0.538, 0.527, 0.563 & 0.526, 0.48, 0.528 & 0.532, 0.489, 0.515 & 0.538, 0.536, 0.589 & 0.537, 0.527, 0.542 & 0.544, 0.56, 0.59 \\
$R_{tb\tau}$ & 1.07 & 1.1 & 1.09 & 1.1 & 1.03 & 1.08 \\ \hline
\end{tabular}}
\caption{Table of benchmark points which exemplify the implications of the model. All points are chosen to be consistent with the constraints given in Eq.(\ref{eq:constraints}) except the Planck measurements. All masses are given in GeV and cross-sections in pb. The relevant masses and parameters are shown in red when they are relevant to the discussions. The parameters leading to inconsistency with the current constraints are displayed in blue.}
\label{tab:BPs}
\end{table}

Before concluding, the observations can be exemplified better by providing a set of benchmark points given in Table \ref{tab:BPs}. These points are chosen to be compatible with the constraints given in Eq.(\ref{eq:constraints}) except the measurements of the Planck satellite on the relic density of LSP neutralino. All the masses are given in GeV and cross-sections in pb. Point 1 represents solutions which yield the lowest fine-tuning measurement. Such solutions usually lead to relatively lighter Higgs boson of mass about 123 GeV. Since this light mass of Higgs boson can be corrected by varying the top quark mass within $1\sigma$, cush light Higgs boson masses are still accepted in the scans. However, if one wants to enhance the SM-like Higgs boson mass to its experimentally measured value without changing the top quark mass, then the fine-tuning is realized relatively greater as exemplified in Point 2. In addition, these two points depict solutions of Bino-like LSP neutralino. As discussed above, the coannihilation or annihilation scenarios cannot be realized in these cases, and consequently they predict very large relic densities for the LSP. The rest of the points display solutions for Higgsino-like LSP. The masses of the particles are highlighted with red when they participate in coannihilation or annihilation processes. All the Higgsino-like LSP solutions exemplify the chargino-neutralino coannihilation scenario. In addition, Point 4 also realizes the stau-neutralino coannihilation and Point 5 shows the $A-$resonance solution in which a pair of LSP annihilates into a CP-odd Higgs boson. Point 5 also depicts a solution of the smallest cross-section for the spin-independent scatterings of Higgsino-like DM, while Point 6 shows the greatest cross-section compatible with the current LZ results. These solutions are expected to be tested soon. Point 6 also represents solutions where the LSP neutralino happens to be Bino-Higgsino mixture, and it reveals that even a small portion of Bino ($\simeq 10\%$) in the LSP composition is enough to strike out the relic density, which is slightly larger than the current Planck bound. It can be concluded that the relic density of LSP increases proportional to the Bino percentage, since it causes to suppression in the coannihilation and/or annihilation processes.

\section{Conclusion}
\label{sec:Conc}

It is discussed the low scale implications of a class of SUSY GUTs confronted with the model independent constraints from a variety of experimental analyses. In the models under concern, the SSB Lagrangian of MSSM is generalized by being extended with the NH terms, which can be induced through the SUSY breaking, while they are forbidden in the non-broken supersymmetric case. Even though it is one of the simplest extension of the usual MSSM construction, the NH terms can considerably alter the implications and bring a significant portion of the fundamental parameter space back to be available. In order to make their effect more visible, the work represented here is limited to a class of SUSY GUTs, which is spurred by restricted boundary conditions imposed at $\mgut$. In addition to a variety of experimental constraints employed in our analyses, it is also imposed the unification of the Yukawa couplings at $\mgut$, which provides a strong impact in shaping the parameter space. In this context, the models discussed here involve those which form a minimally constructed class of Yukawa unified SUSY GUTs called NUHM2 

One of the straightforward effects from the NH terms is to break the direct relation between the Higgsino mass and the fine-tuning. In the absence of NH terms, the Higgsino-like LSP receives a very strong impact from the current DM observations which allows Higgsino LSP solutions heavier than about 1 TeV and such solutions can be realized in highly fine-tuned regions. On the other hand, the NH terms provide extra contributions to the Higgsino mass while they leave the fine-tuning measurements intact. It is observed that Higgsino-like LSP solutions can be found which are consistent with several DM observations while they reside in acceptably low fine-tuning regions. These regions can also accommodate YU compatible solutions. The NH terms can also improve the SM-like Higgs boson mass predictions without moving the supersymmetric mass spectrum to too heavy scales. In addition, it is realized that the YU solutions can also be possible when the SSB mass terms for the MSSM Higgs fields are set universal ($m_{H_{d}}=m_{H_{u}}$) at $\mgut$, which is not possible without non-universalities in the boundary conditions or the presence of NH terms. 

Although they do not affect the stop sector much, the NH terms yield relatively lighter sbottoms in the spectrum whose contributions to the Higgs boson mass become significant in this class of models. The heavy Higgs boson masses receive much larger contributions from NH terms. Depending on the sign of the relevant NH terms, these Higgs bosons can be lighter or heavier, and our analyses result in such Higgs bosons which can be as light as about 600 GeV consistent with the constraints from rare $B-$meson decays.

The DM implications of NUHM2 models in the presence of NH terms can be split into two distinct classes by LSP species. If the LSP is considered to be a candidate, the DM can be formed by Higgsinos, Binos separately, or a mixture of these two particles. The Higgsino-like LSP solutions are mostly favored by the current measurements of Planck satellite by yielding a consistent or low relic density DM solutions. These solutions can lead interesting predictions in different DM scenarios. The relatively lighter mass spectrum also allow several coannihilation/annihilation scenarios in which the relic density of LSP can be significantly reduced to the desired ranges through coannihilation of LSP together with another supersymmetric particle or their annihilations into the non-supersymmetric particles mediated by the heavy Higgs bosons. The chargino-neutralino coannihilation scenario is very typical for the Higgsino-like LSP, but also stau-neutralino coannihilation and $A-$resonance solutions can take part in reducing the relic density of LSP. With the effects from the NH terms, also sbottom-neutralino coannihilation scenario becomes available again, but such solutions are not compatible with the YU conditions, though one can analyze these solutions further without YU.

This work rather emphasizes the implications obtained in our analyzes which cannot be realized in the restricted models when the NH terms are absent such as low fine-tuning, Higgsino-like DM, relatively lighter sbottoms etc. The restricted models leads to quite heavy mass spectrum and Bino-like LSP. Imposing the YU condition gives rise to the masses further, and hence the solutions cannot be accommodated in the regions with low fine-tuning. Such solutions reappear in experimentally consistent regions, because of the relaxation against the phenomenological and indirect constraints arising from the presence of the NH terms. On the other hand, the NH terms cannot help much when one considers more direct and model independent experimental results such as those from the LHC and DM experiments. Despite the lighter Higgs bosons of mass around 600 GeV in the spectrum, $A,H\rightarrow \tau\tau$ events can still exclude the solutions for $m_{A}\lesssim 2$ TeV. A similar impact can be observed in the DM implications. Even though Higgsino-like LSP can be found compatible with the relic density and low fine-tuning condition, the direct detection of DM experiments lead to a strong negative impact on such solutions, since the Higgsino-like DM is predicted to be scattered with a very large cross-section to which the current direct detection experiments are highly sensitive. The Higgsino-like LSP solutions with large cross-section are not currently being excluded, but they are expected to be tested soon in LZ and XENON experiments. Even though negative strong effects may not mean a total exclusion of these solutions, they indicate that the models can need to be extended more than including only the NH terms.

\noindent{\bf Acknowledgment}
The author acknowledges the resources supporting this work in part were provided by the CEAFMC and Universidad de Huelva High Performance Computer (HPC@UHU) located in the Campus Universitario el Carmen and funded by FEDER/MINECO project UNHU-15CE-2848.

\bibliographystyle{JHEP}
\bibliography{QYU.bib}

\end{document}